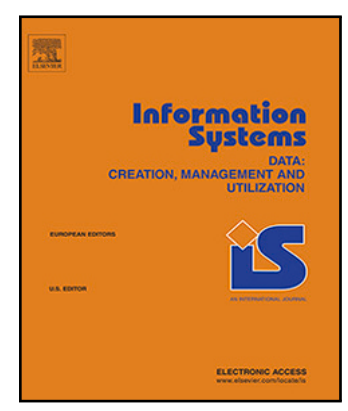

# From precision to perception: Human-in-the-loop evaluation of keyword extraction for internet-scale contextual advertising[☆,☆☆]

Jingwen Cai [a,*], Sara Leckner [b], Johanna Björklund [a]

[a] Department of Computing Science, Umeå University, Umeå, 90187, Sweden

[b] Department of Computer Science and Media Technology, Malmö University, Malmö, 20506, Sweden

ABSTRACT

Keyword extraction is a foundational task in natural language processing, underpinning countless real-world applications. One of these is contextual advertising, where keywords help predict the topical congruence between ads and their surrounding media contexts to enhance advertising effectiveness. Recent advances in artificial intelligence have improved keyword extraction capabilities but also introduced concerns about computational cost. Moreover, although the end-user experience is of vital importance, human evaluation of keyword extraction performances remains under-explored. This study provides a comparative evaluation of prevalent keyword extraction algorithms with different levels of complexity represented by TF-IDF, KeyBERT, and Llama 2. To evaluate their effectiveness, a mixed-methods approach is employed, combining quantitative benchmarking with qualitative assessments from 855 participants through four survey-based experiments. The findings demonstrate that KeyBERT achieves an effective balance between user preferences and computational efficiency, compared to the other algorithms. We observe a clear overall preference for gold-standard keywords, but there is a misalignment between algorithmic benchmark performance and user ratings. This reveals a long-overlooked gap between traditional precision-focused metrics and user-perceived algorithm efficiency. The study underscores the importance of human-in-the-loop evaluation methodologies and proposes analytical tools to facilitate their implementation.

## 1. Introduction

Keyword extraction is a central task in many natural language processing (NLP) algorithms. The objective is to extract a set of words or phrases that can effectively represent a given document. Various algorithmic solutions have been attempted [1–9], differing in how they assign relevance to individual words, and how they combine salient words into a representative set of keywords. From the earliest statistical methods, which focused on word frequencies, to the current applications of large language models (LLMs), keyword extraction has played an important role in various information processing and supporting tasks, such as information retrieval, document categorisation, and sentiment analysis [10,11].

In this article, we examine evaluation techniques for keyword extraction methods, with particular attention to algorithmic aspects that are most relevant to applications in online advertising, where extracted keywords serve as the link between advertisements and user search queries [12] and media contexts [13]. In search advertising, dominating platforms such as Google Ads match user queries to advertiser keywords [14], whereas in contextual advertising, the similarity between the ad and media keywords acts as a proxy for ad–context congruence [13]. Strong congruence has been shown to improve click-through rates for ads placed on news sites [15] and social media platforms [16], and to positively affect brand attitude and ad recall [17,18]. Moreover, keywords play a crucial role in content moderation by helping to avoid negative or sensitive content [19] through mechanisms such as blocklisting [20]. Combined, this leads to improved user experiences, higher advertising efficiency and better return on invested advertising spend [21,22].

Keyword extraction in real-world settings poses several challenges, however. In the case of online advertising, it entails the processing of massive data volumes, and automated bidding protocols require responses within fractions of a second after an ad opportunity appears [23]. Because each auction has low economic value, there are

☆ This article is part of a Special issue entitled: 'HILDA' published in Information Systems.

☆☆ This work was supported by the Marianne and Marcus Wallenberg Foundation, Sweden, the Wallenberg AI, Autonomous Systems and Software Program (WASP) and WARA M&L funded by the Knut and Alice Wallenberg Foundation, Sweden, and the Swedish Research Council.

* Corresponding author.
*E-mail addresses:* jingwen.cai@umu.se (J. Cai), sara.leckner@mau.se (S. Leckner), johanna.bjorklund@umu.se (J. Björklund).

strict limits on preprocessing and computational parallelisation. The uncurated nature of the Internet also introduces substantial noise, demanding robust algorithms. Moreover, practical systems must allow for the detection and mitigation of demographic biases [24], because even without personalised data, media content can indirectly reveal user characteristics. Altogether, these constraints call for keyword extraction methods that are robust, transparent, and efficient.

Considerable effort has been devoted to aligning algorithmic performance with application needs, yet several challenges remain. While the introduction of LLMs has greatly enhanced keyword extraction, it has also raised concerns about computational costs, both in terms of energy use and financial expenditure. Although GPUs enable efficient processing, they require substantial investment. Moreover, different algorithms inevitably generate different keyword candidates for the same content. Even minor improvements in algorithmic precision may require significant research efforts and increased computational resources, yet their impact on users' subjective evaluation is uncertain. In real-world scenarios — where users typically have limited knowledge of the technical principles underlying the information presented — it is unclear whether they would prefer the keywords extracted by more sophisticated algorithms over those generated by simpler, less resource-intensive alternatives. Finally, keyword extraction algorithms are typically evaluated against fixed sets of gold-standard keywords created for general purposes — machine-driven approximations of human judgment intended to enable faster model iteration — rather than through direct user feedback [25,26]. This approach makes the evaluation heavily dependent on the chosen annotations, potentially penalising algorithms that capture underlying meaning rather than exact wording, even when the former better reflects human-perceived coherence [27]. To overcome these difficulties, greater human involvement is needed in algorithm evaluation.

This study introduces a user-centred framework for comparing keyword extraction methods, aiming to identify the defining properties of high-quality keyword sets and to refine reliable assessment tools through a combination of qualitative and quantitative human-in-the-loop evaluation methods. Ultimately, the framework supports researchers and practitioners in choosing more practical and user-preferred keyword extraction algorithms. The study addresses the following research questions:

**RQ1** What factors should be considered when conducting human-in-the-loop evaluations of keyword-extraction algorithms?

**RQ2** How do keywords generated by algorithms with different computing complexities perform in terms of precision-based and perception-based evaluations?

**RQ3** What implications do these findings have for real-world applications and downstream tasks, such as contextual advertising and information retrieval?

To answer these questions, we conduct an experimental study that mainly compares three keyword extraction algorithms representing different levels of complexity: (i) **TF-IDF**, a straightforward statistical method; (ii) **KeyBERT**, a deep neural network for keyword selection; and (iii) **Llama 2**, a robust open-source LLM of medium size. To assess how algorithmic performance affects downstream applications, we conduct a follow-up experiment on document recommendation, which also allows us to evaluate three additional language models.

For the qualitative analysis, user preferences are collected through survey-based user experiments and further analysed through novel analytical tools to manage experimental variables that could otherwise distort user feedback. For quantitative assessments, the keywords produced by the algorithms are benchmarked against participant-defined gold standards using several common metrics. The computational demands of the evaluated algorithms are also considered in the analysis.

The key contributions of this work align with the research questions posed above:

- We propose a human-in-the-loop evaluation framework that combines quantitative precision metrics with qualitative user perceptions to compare keyword extraction approaches.
- We identify defining characteristics of high-quality keywords, design survey-based experiments to capture user feedback, and introduce novel analytical tools to support evaluation.
- Our comparative analysis of keyword extraction methods underscores the importance of human-in-the-loop evaluation and yields insights for real-world applications.

## 2. Theoretical background

Keyword extraction, also known as keyword detection or keyword analysis, involves techniques that automatically extract a descriptive set of words or phrases from a document in a text collection (i.e., a *corpus*). In this context, individual words are often referred to as *unigrams*, e.g., 'horse' or 'paddock', and sequences of more than one word as *N-grams*, e.g., 'uneven surface' or 'famous racing track' [28]. The extracted keywords serve as a compact representation of the document and provide some level of understanding of its contents. Previous studies have shown that a good keyword set should satisfy the criteria of *representativeness* and *distinctiveness* [29]. In [30], *coherence, accuracy* and *coverage* are used to evaluate the retrieved information. These imply that keywords should contribute to an accurate, coherent and comprehensive representation of the document and provide a distinctive characterisation that distinguishes it from others within the same corpus, which inspires the properties used in this study.

### 2.1. Keyword extraction algorithms

While various keyword extraction algorithms have been developed, this study focuses on three fundamental families to explore human-in-the-loop evaluations: conventional statistical methods, modern extensions using deep neural-network-based word embeddings, and zero-shot Large Language Models, each represented by one of their most commonly used members.

One of the oldest, but still effective and commonly used methods for keyword extraction, is TF-IDF [8,31,32]. It measures the relative importance of a word candidate in relation to a single document compared to the background of a larger collection of documents. This method is based on two statistical values: *term frequency* (TF), which counts the number of times a word occurs in a document, and *document frequency* (DF), which counts the number of documents in a collection in which the word occurs. It is assumed that the higher the term frequency, the greater the probability that a word is important for a document. By considering the inverse document frequency, the interference from high-frequency words in the general domain (e.g., 'the' and 'are') can be eliminated. When both TF and IDF are high, then the word is considered a good representative that helps distinguish the document from the rest of the collection. Over the years, the method has been used successfully in various domains [33,34]. In an advertising context, fewer examples exist, but it has been used to improve topical advertisement matching using learning-based techniques [35] and to combine semantic and syntactic features for advertisement matching [36]. However, although TF-IDF is easy to implement and computationally efficient, it lacks the ability to distinguish between the different semantic meanings of a term. For example, it may conflate polysemous terms such as 'bank', referring to both (a) a land alongside a body of water and (b) a financial establishment, into single entities. This limits the accuracy of the method.

The second keyword extraction method used in this study is KeyBERT, developed by [37]. As the name suggests, KeyBERT builds on the pre-trained language model Bidirectional Encoder Representations from Transformers (BERT) [2]. It is a simple and user-friendly technique that uses BERT embeddings to generate keywords and keyphrases most similar to a document, and is, compared to BERT, specifically tailored for keyword and phrase extraction. Unlike TF-IDF, KeyBERT captures the semantics of the words, translates them into *embeddings*, that is, representative vectors, and uses these as the basis for selecting keyword candidates. KeyBERT is suitable for a variety of tasks. However, it is more computationally intensive than TF-IDF, as it relies on a neural network for the central part of the computation. Previous studies comparing keyword extractions using standard techniques such as TF-IDF with KeyBERT have shown that KeyBERT performs better in terms of higher similarity to author-assigned keywords [38] and on pre-trained domain-specific datasets [39]. A key reason for this improvement is that KeyBERT takes into account the semantics and the context of the given text, which statistical methods such as TF-IDF, currently ignore [38]. However, the performance of KeyBERT has been shown to vary depending on the specific applied domain [29,40] and the fine-tuning conditions [39].

Since KeyBERT was developed in 2020, approaches have increasingly shifted from extraction to generation, as neural sequence-to-sequence models and LLMs are able to produce more flexible and semantically meaningful phrases. This shift is reflected in recent advances, such as domain adaptation techniques that take advantage of citation contexts (e.g., SILK, [41]), frameworks for on-demand generation guided by user intent (e.g., MetaKP, [42]) and hybrid approaches that extend traditional extraction methods by integrating LLMs, for instance, in KeyLLM, an extension of KeyBERT [43]. KeyBERT, as well as TF-IDF are, however, still widely used [44–46].

The third extraction method to be evaluated is the LLMs, represented by Llama 2. In February 2023, Meta introduced Llama as a pre-trained, publicly available, large language model with parameters between 7 billion and 65 billion [47]. Based on the Transformer architecture, Llama was trained using only public data to achieve better performance with an open source [48]. Although there are other open-source LLMs, such as open pre-trained Transformer language models [49] and GPT-NeoX-20B [50], Llama shows the ability to compete with proprietary LLMs in certain cases. Most Llama-13B performances outperform GPT-3, even with ten times fewer parameters [48]. Five months after its initial launch, Llama 2 was released, with parameters scaling from 7 billion to 70, increasing the pre-training data by 40% [51]. This update also compensates for the lack of alignment optimisation in the initial model and launches a fine-tuned chat model using reinforcement learning with human feedback (RLHF) and other fine-tuning techniques [52]. In addition to performance improvements over Llama, Llama 2 is now available for various markets, expanding its applications beyond research and opening up more opportunities for commercial use. Since the introduction of Llama 2, the development of LLMs has progressed towards increased computational efficiency, openness, and multimodal integration. Notable developments in this context are ChatGPT 4.0, Llama 3, Mistral, and DeepSeek.

There is a growing amount of literature that focuses on the automated evaluation of LLMs across a spectrum of NLP tasks, for example, question-answering and sentiment analysis. For instance, in [53], the authors compared the performance of ChatGPT with other recent LLMs, including GPT-3.5 and PaLM, across a range of NLP tasks. The evaluation was based primarily on publicly available datasets, with accuracy used as the main performance metric. In contrast to this study, their aim was to investigate whether ChatGPT exhibits variations in performance across different tasks, without collecting user ratings or subjective perceptions of these performances.

### 2.2. Human versus algorithmic evaluation

Evaluations of keyword extraction algorithms typically rely on automated methods [26,27] and focus on precision and recall against pre-determined keywords, considered the gold standard [25]. This approach enables comparability across studies using the same dataset, but may not fully capture how well algorithms meet human expectations and needs. Since humans are usually the end users of these algorithms or the downstream NLP tasks, user-centred evaluation plays a crucial role. Humans are generally better at understanding fundamental aspects that automated methods often struggle with, such as semantics and context [54], although the introduction of LLMs has narrowed this gap to some extent. Although the impressive performance of GPT-3, when first introduced with 175 billion parameters, may be largely attributed to its sheer scale, its subsequent version, ChatGPT, shows that the fine-tuning of domain-specific data combined with RLHF, is just as effective, if not more so, than solely relying on parameter count. These advancements not only demonstrate the efficient use of vast user data, but also highlight the interconnectedness between AI technology and real-world applications. As a result, we are witnessing the emergence of growing subfields of machine learning that are primarily concerned with instruction tuning and human-in-the-loop approaches. For example, [30] used human preference data to improve a model in the style of GPT-3. Compared to this study, they focused on the summarisation task and how performance can be improved through continued training with human feedback. They used the evaluation criteria *coverage* (importance of information), *accuracy* (fidelity to the original information), *coherence* (readability), and *overall quality* for their user evaluation. In another relevant study, [55] trained an information-seeking dialogue agent to be more helpful, correct, and harmless using human feedback. Additionally, [56] evaluated the performance of Llama and other pre-trained LLMs to demonstrate how they fine-tuned to extract useful records of complex corpora, using a human-in-the-loop annotation process. Compared to this study, human annotations were only used to train the LLMs.

While many studies emphasise the importance of incorporating human preferences into model training, few explicitly analyse user evaluations. Despite advances in algorithmic techniques, comprehensive user evaluations remain under-utilised and under-studied as a means of performance assessment. Nonetheless, a growing body of research has assessed the effectiveness of information retrieval systems [57–59], topic model interpretability [27] and cluster sample validation [60], based on human judgements, primarily through surveys and open labelling. These areas face challenges similar to those in evaluating keyword extraction algorithms, such as this study, often prioritising system metrics over user-centred perspectives. This oversight stems from the practical challenges of manual evaluation [61,62]. First, human evaluation is resource-intensive, time-consuming, and often non-replicable due to the size and domain specificity of the corpora involved [63]. Second, results vary with the evaluator's level of domain expertise. For instance, a layperson assessing the content intended for specialists may yield different outcomes than domain experts [64]. Third, human evaluation is inherently subjective, requiring reliable interpretations, and can lead to inconsistencies across different evaluators or even with respect to the same evaluator over time [65]. To address these limitations, involving a larger pool of non-expert evaluators (which has been shown to produce comparable results to expert assessments [59,62]) and comparing user-centred and machine-based evaluations [66] are recommended strategies for enhancing reliability and efficiency.

Overall, studies focusing on human evaluation of keyword extraction are limited. However, in real-world applications where relevance and user engagement are critical, there is a clear need to assess these methods based on user experience. Integrating human-in-the-loop approaches is essential not only for improving extraction algorithms but also for developing and refining evaluation methods themselves.

### 2.3. Contextual advertising

Contextual advertising involves the placement of ads in 'relevant' media contexts [67,68], a practice that often depends on high-quality targeting keywords. The notion of relevance is multifaceted, which can mean a topical congruence between the media contexts and the ad (e.g., both are about electric cars), or that the context primes for favourable associations with the product (e.g., the media context is about plane crashes, and the ad is about city jeeps, which is a comparatively safe car model). Previous research has demonstrated that the perceived relevance (i.e., personal context of meaning) has a key role in favourably impacting the effectiveness of advertising messages on cognitive, affective, and behavioural levels [21,22]. Extant findings suggest that users are more inclined to interact with advertisements that are thematically congruent with their surrounding editorial content [69,70]. In this case, the user views the ad and the webpage as value-adding entities, which positively influence each other, rather than being perceived as intrusive and disturbing [67]. Given the escalating debate around privacy and the — actual or perceived — illegitimate use of personal data, contextual advertising is offering an attractive alternative to online advertising. This increases the value of high-quality keyword extraction, and makes the choice of targeting keywords a matter of great importance.

A common means of placing ads contextually is to compile a set of target keywords, and compare them to a set of keywords extracted from a candidate webpage, such as a news article in an online newspaper. For example, suitable contexts for a car ad might include words such as 'four cylinders', 'top gear', and 'auto shops'. Advertisers can also try to avoid negative words such as 'collision', 'cannabis', or 'injury'. If the sets of the desired and extracted keywords are sufficiently similar, advertisers bid to place their ads on those webpages. Since the automated auctions used to sell online advertising space must be completed within approximately 100 ms [23], the algorithms involved must operate swiftly and efficiently. Additionally, given the high volume of ad placements traded every second, which can reach hundreds of thousands even on small platforms, the algorithms must be cost-effective. Therefore, when evaluating these keyword extraction algorithms, it is important to take computational cost into account, along with human-in-the-loop evaluations. This underscores the potential for further research to incorporate user-centred methods, and provides valuable insights into the effectiveness and user satisfaction of emerging algorithmic approaches.

## 3. Methodology

As mentioned in the previous section, user annotations are commonly used to create benchmark datasets and optimise algorithm performances. However, they are less frequently used for direct evaluations of algorithms, which overlooks the fundamental user experiences in the practical sense. While the technical performance of algorithms can be quantitatively assessed by comparing algorithmic outputs to benchmarks, real-world applications such as advertising often depend on user experiences and perceptions. In such cases, a five-per-cent increase in algorithm precision may yield different targeting keywords, but does not necessarily lead to a proportional increase in user click-through rates. Therefore, this study incorporates classic precision-based quantitative benchmarking with a perception-based, qualitative evaluation of algorithm-generated keywords, which consists of four survey-based experiments on crowdsourced participants.[1] In addition, the experimental design takes into account the practical challenges associated with human evaluation, as discussed in Section 2.

To address RQ1 and prepare for user evaluation, we define a set of keyword evaluation properties introduced in the next section. For RQ2, we quantify the standard quality metrics across the evaluated algorithms and analyse the experimental results in relation to the defined keyword properties. Finally, we address RQ3 by further analysing the experimental results in light of user-centred evaluation.

### 3.1. Target properties

Before conducting the experiments on the crowdsourced participants, a set of properties characterising high-quality keywords was identified. They were inspired by [29,30], and tailored to capture fundamental aspects of human evaluations, particularly in the context of keyword extraction tasks.

The selection of comprehensiveness, representativeness, distinctiveness, and reasonableness as evaluation metrics follows from their complementary roles in ensuring both coverage and quality. *Comprehensiveness* captures the extent to which all relevant aspects of the domain are considered, while *representativeness* ensures that the extracted elements adequately reflect the broader population or corpus. *Distinctiveness* is crucial for identifying features that uniquely differentiate one set of items from others, thereby avoiding triviality and highlighting salient themes. Finally, *reasonableness* guarantees coherence, appropriateness, and semantic accuracy, ensuring that the results are interpretable and not merely idiosyncratic. Taken together, these properties jointly provide a balanced framework for evaluating the relevance and usefulness of extracted keywords.

Given document $d$ and an ordered set $W = \{w_1, w_2, \dots\}$ of keywords extracted from $d$, we say that the set $W$ is a *proper* keyword set for the document $d$ if it is:

- ***comprehensive*** in that it covers multiple aspects of $d$,
- ***representative*** in that it captures the central features of $d$,
- ***distinctive*** in that its keywords stand out as unique for $d$ compared to the other words in $d$, and
- ***reasonable*** if the keywords appear to be chosen according to coherence, reasonableness, and principles.

As we will further discuss in Section 3.5, these properties are partially overlapping, but combined we argue that they provide a fair predictor of overall quality. To validate this claim, participants were asked to give explicit feedback on the subjective *overall quality* of the presented keyword sets. Moreover, during the participant evaluation, gold-standard keyword sets were collected (see Section 3.5.1) and used to quantitatively calculate standard quality metrics in terms of precision, recall, cosine similarity, and edit distance (see Section 4.6). *Precision* is the ratio of the algorithmically extracted keywords that also occur in the gold standard relative to all extracted keywords; *recall* denotes the ratio of gold-standard keywords that appear in the algorithmically extracted set; ***cosine similarity*** shows the distance between two keyword sets in the vector space; and *edit distance* measures the differences between extracted keywords and the gold standard as the smallest number of operations required to rewrite the former into the latter.[2] Ideally, keywords extracted by a good algorithm should achieve high precision, recall and cosine similarity with a small edit distance, compared to the gold standard.

### 3.2. Dataset

A primary target of contextual advertising is online news, due to the quality and volume of Internet traffic it generates. In addition, its literary style lends itself to a suitable genre for algorithmic analysis. While other forms, such as tweets and videos, are also widely used and potentially more profitable, they involve additional complexities and variables like sentiment analysis and multimodal techniques. Therefore, news articles were selected for the experiments in this study.

The articles making up our dataset were accessed through Aeterna Labs.[3] This dataset comprises over 45 000 news articles from the British newspaper *The Guardian*, most of which were published in 2022. First, the dataset was filtered to remove articles that failed to meet the criteria for accessibility and quality, or that exceeded 350 words in length, resulting in a candidate pool of 5986 news articles. From this pool, five articles were manually selected from a random sample of 50 to ensure topical diversity — society, sports, culture, music, and wildlife — while ensuring the use of accessible language without domain-specific terms. These constituted the final set of **base articles** used in the experiments (see Appendix A, Table A.1). For each base article, the keyword extraction algorithms TF-IDF, KeyBERT, and Llama 2 were applied to generate ten keywords per article (see Appendix A, Table A.2). The resulting $5 \times 3$ combinations formed the initial task instance pool for the experiments. In addition, a 'gold-standard' keyword set was created for each base article based on participant annotations (see Section 3.5.1).

[1] Research data is available for download at github.com/JingWen17/Precision_to_Perception.

[2] The evaluation in terms of the edit distance is also the reason why we chose to work with ordered sets as opposed to regular, unordered sets, because the measure is only applicable if the elements are ordered. As both the algorithmic extraction methods and the way we generated gold-standard keyword sets from manually annotated data suggest an ordering of the keywords, this is not a limitation.

[3] Aeterna Labs is a Swedish adtech company; see https://aeternalabs.ai/.

Widely used benchmark datasets for keyword extraction are often small or originate from genres other than real online news, making them less relevant to the purposes of this study. While exceptions like *KPTimes* [71] and *Mopsi Newspaper Dataset* [72] exist, they are slightly outdated compared to our dataset. Most importantly, in this study, both the gold-standard keyword selection and rating tasks are conducted by participants recruited from the same crowdsourcing platform under the same screening conditions, ensuring consistent standards and minimising potential variables compared to using existing benchmark gold standards selected by external, unknown annotators. From here on, for clarity, we define *base articles* as these selected five news articles, and *task instances* refer to the combinations of these base articles with different experimental manipulations, such as keywords extracted by different methods.

### 3.3. Technical implementation

Data pre-processing used the *ROUGE* list [73] to remove stop words, but was otherwise kept to a minimum to avoid influencing the keyword extraction algorithm comparisons. Keywords were extracted using the default 1-gram settings to adhere to the minimal-variable and standard implementation approach. The pre-trained model `all-MiniLM-L6-v2` [37] was used for KeyBERT and the 7 billion parameter version for Llama 2. The choice of the medium-size Llama 2 model was motivated by the scale of the datasets involved in contextual targeting (see Section 2.3). Moreover, as the 7B model was compact enough to be hosted on a local machine, uploading copyrighted content to the cloud could be avoided. To extract keywords with Llama 2, the prompt scheme from the KeyLLM project[4] was used, without explicit instructions on the specific properties.

### 3.4. Experimental setup

Participants for the experiments were recruited via the *Prolific* platform.[5] Prolific enables participant selection based on segmentation criteria such as demographics and consuming behaviour, ensuring that the sample matches experimental objectives. A total of 855 participants (see Appendix B for demographics) residing in the UK were recruited, to reflect the task articles' origin. Before starting the tasks, the participants were provided with a brief introduction to the study, including its general purpose and ethical considerations. To ensure anonymity, no identifying information was collected. Only responses from participants who completed the entire task were considered valid and included in the final analysis.

The study consisted of four survey-based experiments. The surveys used in each experiment were administered through an online form, created using the web-based annotation tool *Potato*[6] [74]. Each survey was structured into four sections. The first introduced the task, obtained consent for participation, and provided task instructions. The second collected basic demographic information. Next, in the main section, each participant was randomly assigned several task instances for assessment. Each assessment involved reviewing the associated task instance and rating it on a seven-point scale using Likert-type questions (e.g., see Section 3.5.1). Upon completing all assigned task instances, in the last section, participants were invited to provide feedback on their experiences with the experiment.

### 3.5. Keyword-extraction experiments

This section outlines the implementation details of the experiments conducted in this study, each following the setup introduced in Section 3.4 above. The *preliminary experiment* aimed to test the interactive workflow, decide between alternative ways of phrasing the task questions, and verify technical functionality for the *main experiment*, which was then conducted to collect evaluation data on the performance of each extraction algorithm in isolation (i.e., using the default task instance only showing a task article with the keywords generated by one of the methods). To support further analysis of user perceptions, a *supplementary experiment* was then performed to allow a direct comparison between the keyword sets generated by all four methods. All of these three experiments aimed to explore how participants perceive keywords extracted from a given base article using Gold-Standard, TF-IDF, KeyBERT and Llama 2. In contrast, the *downstream experiment* simulated real-world contextual recommendation tasks by identifying the most relevant news articles for each base article based on keyword similarity, using keywords extracted by an expanded set of methods.

#### 3.5.1. Preliminary experiment

A key aspect of the preliminary experiment was determining the phrasing of the questions (used in the main experiment) to ensure clarity and reduce the risks of poorly worded questions affecting the quality and independence of participant responses. Accordingly, each of the five properties described in Section 3.1 was assessed using three differently formulated questions:

1. ***Comprehensiveness***
   (a) How well do the keywords cover the main article?
   (b) To what extent do you think these keywords cover important information from the article?
   (c) To what degree do you think these keywords comprehensively summarise the main article?
2. ***Representativeness***
   (a) To what extent do the keywords adequately capture the properties of the main article?
   (b) In your opinion, how accurately do these keywords represent the overall article?
   (c) To what degree do you think these keywords represent the original article?
3. ***Distinctiveness***
   (a) How unique do you find these keywords in relation to the original article?
   (b) How do these keywords stand out when compared to other words in the article?
   (c) How would you rate these keywords in terms of their distinctiveness for the article?
4. ***Reasonableness***
   (a) In your opinion, does the selection of these keywords seem reasonable in the context of the article?

[4] See https://maartengr.github.io/KeyBERT/guides/keyllm.html.
[5] See https://www.prolific.com.
[6] See https://potato-annotation.readthedocs.io/en/latest/.

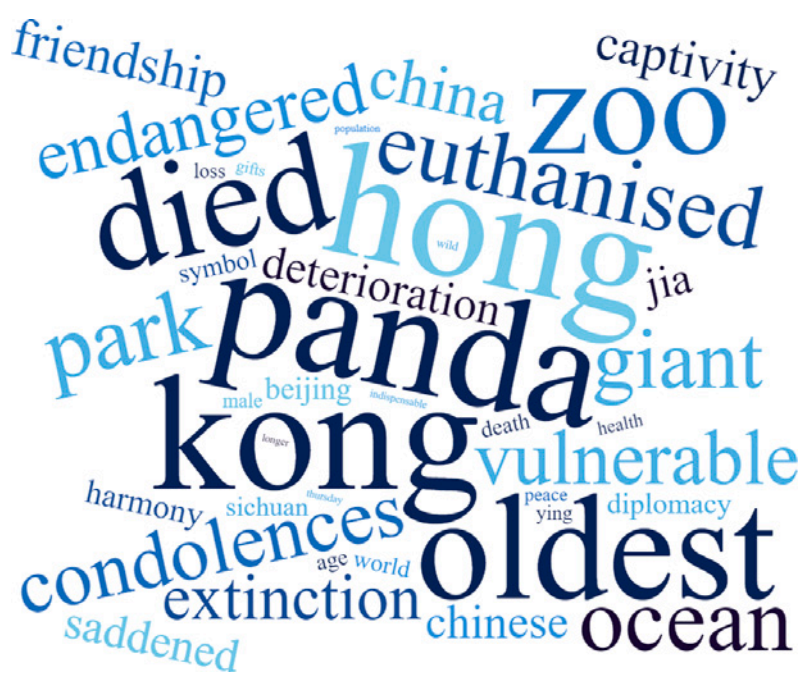

**Fig. 1.** A word cloud created from participants' keyword selections for a base article: Word size reflects selection frequency with larger words indicating higher participant agreements. Words with a frequency of less than five are excluded from this figure.

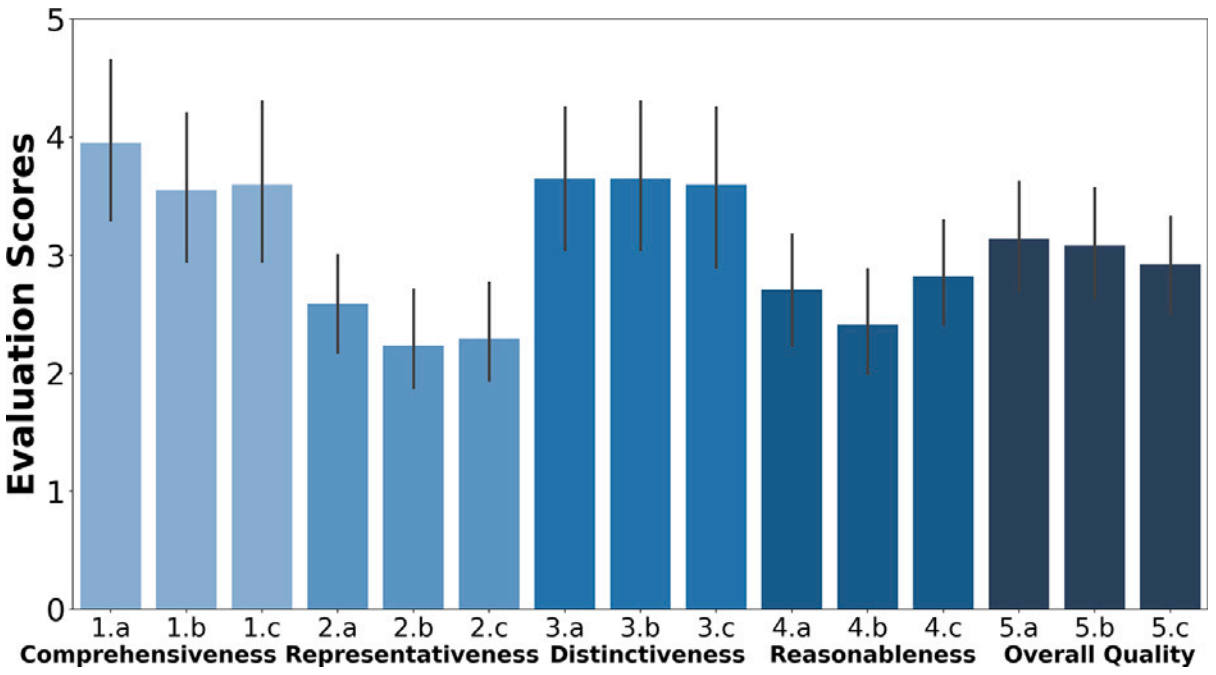

**Fig. 2.** Example of horizontal analysis of feedback consistency: The analysis shows participants' responses for a specific task instance, with 20 participants assigned to this instance and question group. The $x$-axis represents alternative question phrasings used in the preliminary experiment, while the $y$-axis shows the mean participant scores. Black lines indicate standard deviations.

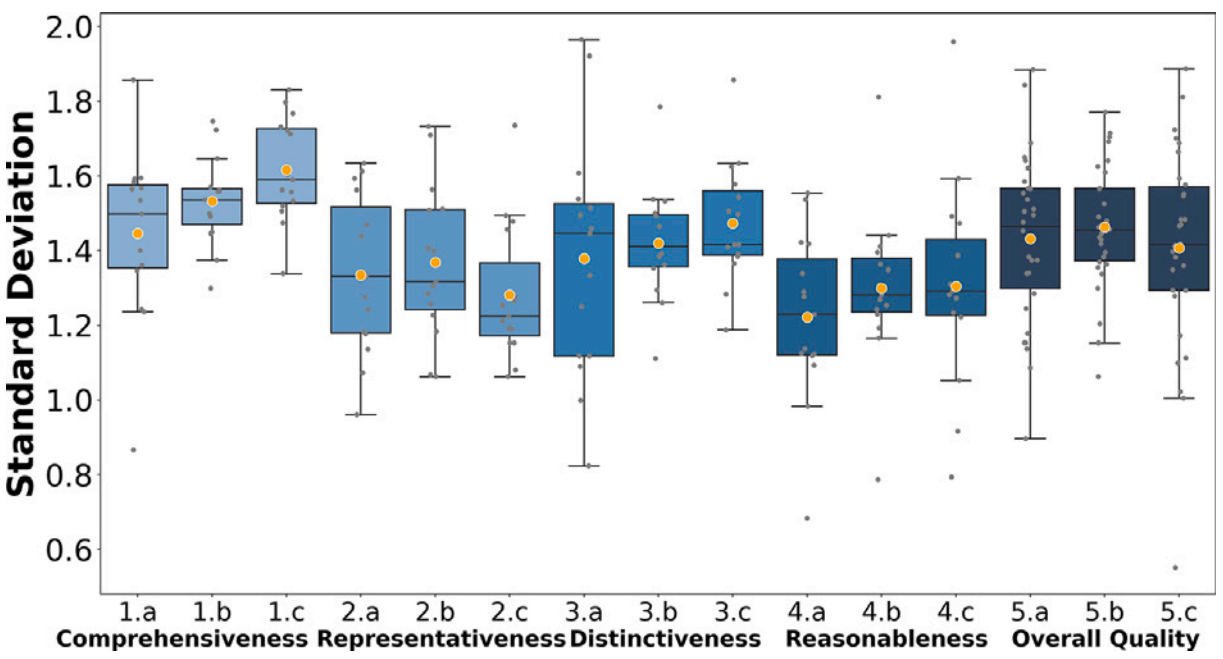

**Fig. 3.** Distribution of each question score's standard deviation under horizontal analysis averaged over *all* task instances (in contrast to Fig. 2, which presents one task instance): The $x$-axis shows alternative question phrasings; the $y$-axis displays each instance-level standard deviation with a grey dot, and orange dots indicate the mean standard deviation per phrasing across all task instances. Box plots show the distribution of standard deviations: the upper and lower lines indicate the upper and lower quantiles, with the inner horizontal line marking the median. (For interpretation of the references to colour in this figure legend, the reader is referred to the web version of this article.)

(b) If you were asked to select a set of keywords from the article, then how would you rate the reasonableness of the given ones compared to your selection?
(c) To what degree do you think these keywords make sense in relation to the original article?

5. ***Overall quality***

(a) Considering the previous questions, what is your general opinion of these keywords for this article?
(b) To what extent do you agree with the selection of these keywords based on their overall quality to reproduce the original article?
(c) How would you evaluate the overall quality of these automatically generated keywords?

In addition, during this experiment, participants were asked to select ten keywords from each task article based on their personal judgments. However, the number was not strictly enforced, allowing some participants to choose fewer or more than ten keywords per article. This flexibility aimed to reduce task fatigue, which could otherwise lead to random word selection. Based on the collected participant annotations, gold-standard keyword sets were created with the ten most frequently chosen words for each article. Fig. 1 shows a sample word cloud illustrating the words and their selection frequencies from participant annotations for a base article (see Table A.2 in Appendix A for all gold-standard sets).

The preliminary experiment involved a total of 196 valid participants. Each participant was randomly assigned three task instances covering different base articles, resulting in approximately 40 data points for each of the 15 base articles and keyword set combinations from the initial task pool. To reduce the mental load of answering the above-defined 15 task questions for each of the three assigned task instances, the questions were split into two sets: *(comprehensiveness, distinctiveness, overall quality)* and *(representativeness, reasonableness, overall quality)*. Therefore, each task instance assessment comprised only one question set, requiring participants to complete nine Likert ratings and select ten keywords. Fig. E.1 in Appendix E shows the structure of the task page. After rating all three task instances, participants were asked about their fatigue levels. This feedback helped calibrate the number of task instances assigned to each participant, balancing engagement with annotation efficiency (see Appendix C for the query used).

To understand the impact of alternative evaluation question phrasings, two novel types of statistical analyses were developed: *horizontal* and *vertical* analysis. In the horizontal analysis, one combination of task instance and property $c$ was considered at a time. For each of these combinations, the consistency of the participants' answers was measured to three different phrasings that asked about the same property $c$. For example, the middle three columns in Fig. 2 show the average scores and standard deviations of the three phrasings on distinctiveness. In this case, participants' scores appear to be fairly independent of the choice of phrasing. It makes little difference whether question 3.a, 3.b, or 3.c is used, $\chi^2(df = 38, N = 60) = 3.32$, $p > .99$. Fig. 3 summarises the proportion of the standard deviations of all 15 task instances in the horizontal analysis. For each task instance and question, its *dispersion* was assessed by calculating standard deviations of the participant ratings collected. Intuitively, a phrasing is considered effective if there is a high consensus (i.e., low standard deviations) among the participant ratings on the same task instance, and a consistently low mean standard deviation suggests strong overall agreement. Across all 15 task instances, the least consistent phrasing resulted in the largest total standard deviations and should therefore be avoided in the main experiment.

In the vertical analysis, variation in individual participants' responses was examined: When an individual evaluates the same set of keywords under the same evaluation property with different phrasings, the average score is assumed to reflect their 'true' opinion. Correspondingly, a question phrasing is considered effective if participant ratings are fairly stable with respect to this average score. Thus, for all task instances, a smaller difference between a specific phrasing's score and the average score across all phrasings under the same property indicates that the phrasing is more indicative. Fig. 4 shows an example of the scores and absolute deviations of one question set answered by a participant. Therefore, the most consistent property phrasing had the lowest sum of absolute deviations over all instances. Fig. 5 further summarises the proportions of these absolute deviations in the vertical analysis across all participants.

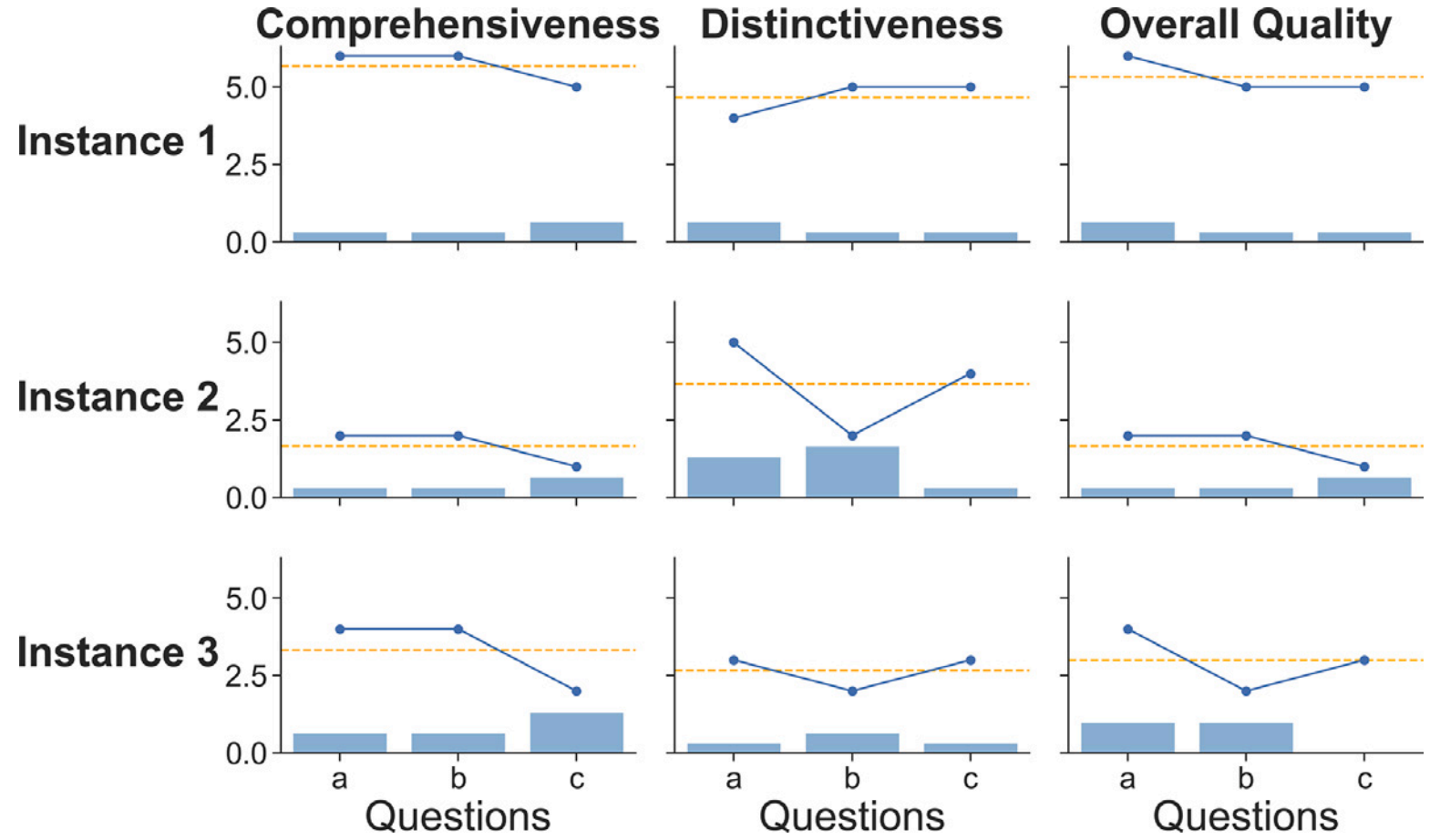

**Fig. 4.** Example of vertical analysis of feedback consistency of a single participant's responses across three assigned task instances: Each column represents a property and each row corresponds to a task instance. The $x$-axis shows alternative phrasings for each property; the $y$-axis shows the participant's score for each phrasing-instance combination. Horizontal orange lines indicate the mean scores across all phrasings for the same property and task instance. Bars represent the absolute value of the difference between a phrasing score and the corresponding mean score. (For interpretation of the references to colour in this figure legend, the reader is referred to the web version of this article.)

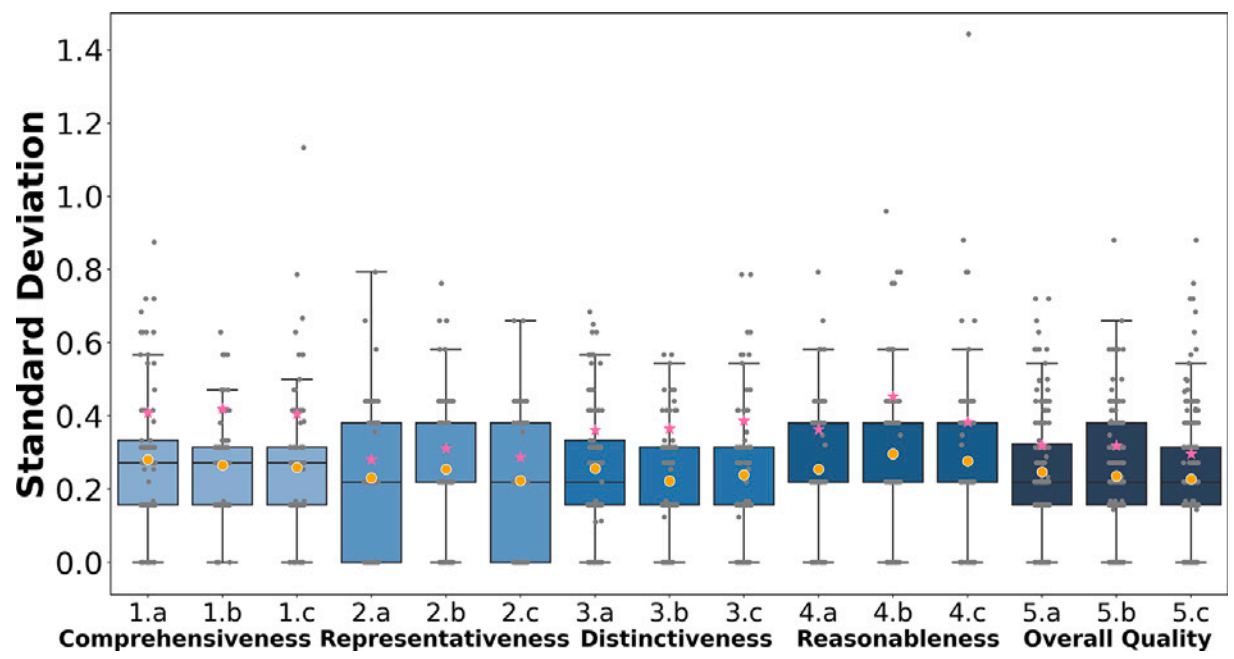

**Fig. 5.** Distribution of each question score's standard deviation in vertical analysis across all participants: The $x$-axis represents different phrasings of the key properties. Pink stars indicate the mean of these differences (visualised as bars in Fig. 4) across all participants. Each grey dot represents the participant-level standard deviation of these differences. The average standard deviations are shown as orange dots. Box plots show the distribution of standard deviations using the same visualisation style as in Fig. 3. (For interpretation of the references to colour in this figure legend, the reader is referred to the web version of this article.)

### 3.5.2. Main experiment

Based on the horizontal and vertical analyses in Section 3.5.1, questions 1.a, 2.c, 3.a, 4.a, and 5.c were chosen to represent the $5 \times 3$ question battery (see Appendix F for details), and used as task questions in the main experiment. Limiting one question per property aimed to make the task more manageable for the participants. In addition to the automatically extracted keyword sets, the gold-standard keyword sets, stemming from participant annotations in the preliminary experiment, were also included without disclosing the fact that they were not algorithmically generated.

In the main experiment, each participant was randomly assigned four task instances related to different articles (see Fig. E.2 in Appendix E for the design of the task instance page used), compared to the three in the preliminary experiment. This was based on the reduced number of task questions per instance, and on the levels of fatigue reported in the preliminary experiment. Additionally, the number of evaluation data points required for each of the 20 task instances was increased to 50, compared to the preliminary experiment, where the focus was on preparation. Thus, a total of 264 participants were included in the main experiment.

### 3.5.3. Supplementary experiment

In both the preliminary and main experiments, each task instance involved assessing a single set of keywords in relation to the task article from which they were extracted, minimising the number of independent variables that could influence participants' responses. To facilitate a direct comparison between keyword extraction methods, a supplementary experiment was conducted in which the participants ranked alternative keyword sets extracted from the same article. In other words, each task instance now contained one base article with all four keyword sets — those generated by TF-IDF, KeyBERT, Llama 2, and the gold standard. In addition, instead of Likert-based questions, participants were asked to choose the best or worst keyword set over the properties *comprehensiveness*, *representativeness*, *distinctiveness*, and *reasonableness*, using similar phrasings in the main experiment. Then the participants were asked to rank the four keyword sets with respect to their *overall quality*. Therefore, each task instance assessment required the participants to answer eight single-choice questions and one sequencing question (see Fig. E.3 in Appendix E for the details of this task instance page).

A total of 92 participants were included in this experiment, ensuring at least 50 evaluation data points per base article. Each participant was randomly assigned three task instances, which were reduced to mitigate potential fatigue due to the expanded set of questions. In each instance, four keyword sets were randomised and presented as *Group A, Group B, Group C,* and *Group D* to prevent participants from perceiving a particular order among the keyword groups that might influence their evaluations, such as the perception that the sets were ranked from least to most sophisticated.

## 3.6. Document-recommendation experiment

To simulate real-world recommendation tasks that rely on keyword extraction, we conducted a downstream experiment in which participants assessed the relevance and topical similarity between pairs of news articles. The article pairs were matched by identifying the five most relevant news articles from the candidate pool for each of the base articles, based on keyword similarity. To provide a broader comparison, an expanded set of keyword extraction methods was employed. In addition to TF-IDF, KeyBERT, and Llama 2, we included Llama 3.1 (8B parameters), Mistral (small) and DeepSeek (v3.2-Exp), all following the same keyword extraction prompting scheme used for Llama 2 in the aforementioned keyword experiments. For efficiency, we used the APIs of these models to extract keywords for all 5986 articles in the candidate pool. The entire extraction per model involved approximately 2.4 million input and output tokens, incurring costs of $0.30 for the Llama 2 API hosted by Replicate,[7] $0.32 for Mistral, and $0.61 for DeepSeek. The Llama 3 API hosted by the Wallenberg Research Arena for Media and Language[8] was accessed on a research request without fees.

After keyword extraction, a Word2Vec model trained on all articles in the candidate pool was used to compute the similarity between the keywords of each base article and the keywords of each candidate article, using the same extraction method. Finally, for each method and base article, the five most relevant articles were selected and paired with the base article. This results in a total of $5 \times 5 \times 6$ (base article $\times$ matched article $\times$ method) task instances. Similar to the other keyword experiments, participants were asked to rate the pair of articles through a series of 7-point Likert scale questions, which were designed to measure their perceived contextual relevance and topical similarity. The survey questions are detailed in Appendix D.

Valid responses from 303 participants were included in the analysis of this experiment, ensuring at least 10 evaluation date points per task instance and at least 50 date points per combination of base article and method. Each participant was randomly assigned five base-matched article pairs from the total of 150 task instances, covering all base articles. In other words, each participant reviewed all base articles once, each paired with a random one of the five matched articles generated by a random one of the six methods. Table E.1 in Appendix E compares the numbers of task instances assigned across the four experiments. Fig. E.4 in Appendix E illustrates the task instance interface designed for this experiment.

## 4. Results and discussion

This section presents the results of the four experiments described in Section 3. Subsequently, the outcomes are further analysed, followed by an analysis and discussion of emerging patterns. We begin by evaluating the soundness of the overall approach.

### 4.1. User evaluations as an indicator of algorithm performance

Following recommendations from prior studies [66], we use a combined manual and automated evaluation approach. To reduce the cost and time of manual human assessment [59,62,63], a larger number of non-experts are engaged in the evaluations. Concerns regarding subjectivity [63,65] are addressed through a systematically designed series of survey-based experiments. To analyse the effects of different question phrasings for each property, we introduce the *horizontal* and *vertical* approaches (Section 3.5.1). Both rely on the assumption that, for a given task instance or participant, the standard deviation of responses across alternative phrasings of the same evaluation property should remain consistent. Inconsistent phrasing suggests low reliability in participant feedback and should be excluded from the main experiment. While *Kendall's W* [75] is a commonly used nonparametric statistic to test assessment consistency and inter-rater reliability often for ordinal ratings, its general use requires that each task instance must be evaluated the same number of times. This condition is not met in our experiments, where we only ensure that each task instance receives more than the baseline number of valid data points defined for the final evaluation; the remainder of the sampling procedure is entirely random. For this reason, we do not apply *Kendall's W* in our analysis.

[7] https://replicate.com/blog/run-llama-2-with-an-api.

[8] http://hub.waraml.org.

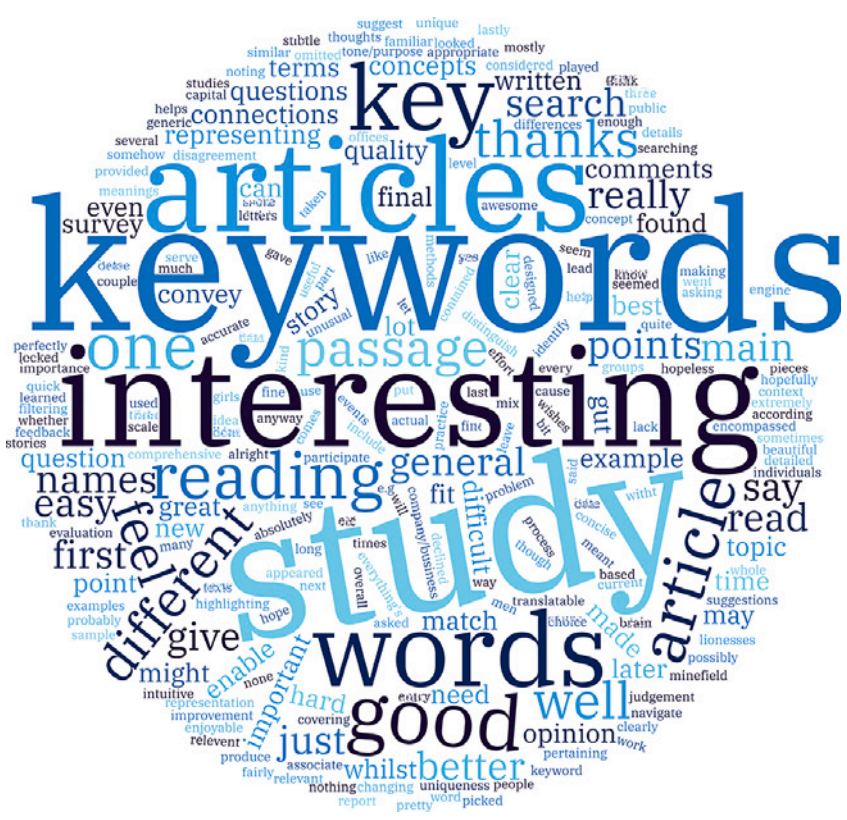

**Fig. 6.** A word cloud generated from participant comments across the keyword experiments: Word size reflects frequency, with larger words indicating more frequent comments.

The preliminary experiment confirms the feasibility of the experimental workflow. Real-time monitoring, enabled by the connection between the experiment's survey interface and the local computer, allows timely resolution of any issues. In addition, the open-ended feedback boxes in the survey offer participants the opportunity to share their experiences (Fig. 6), providing both validation of the setup and a deeper understanding of participant perceptions — a strategic way shown to enhance quality and interpretation [76]. Although task instances are randomised, none are disproportionately resampled by chance, resulting in a relatively uniform number of evaluations per instance. With a sufficient number of participants recruited, we ensure that the total number of evaluation data points for each task instance matches our pre-determined baseline. Moreover, of the 182 participants who comment on fatigue level, 142 respond that three task instances are manageable (see Table C.1 in Appendix C), confirming that the assigned workload is reasonable. These form the basis of the main and supplementary experimental designs and help to reduce the potential of participants providing meaningless feedback due to fatigue and boredom. Maintaining a manageable task load is critical for validity, as overly large test sets can harm participant engagement and their rating qualities [57,77]. The increased number of assigned task instances in the downstream experiment was primarily due to the expanded total of 150 task instances. To ensure sufficient valid data for robust analysis, each participant was required to assess five task instances. However, we ensured that each participant reviewed different base articles with their corresponding matched articles, reducing task repetition. Participants were compensated at a rate of at least $8.00 per hour, in accordance with the standards of Prolific.

### 4.2. Target properties used for keyword evaluation

RQ1 aims to evaluate keyword sets regarding the properties of *comprehensiveness*, *representativeness*, *distinctiveness*, and *reasonableness*, tailored to capture key aspects of how humans assess keyword quality. The results of three keyword experiments confirm that these properties serve their intended purpose, with the exception of *distinctiveness*, which receives notably low ratings from participants (see Figs. 7–9). Open-ended responses indicate that while the keywords capture the general information of the articles, they often miss specific details, such as individual names and the nuanced tone of the article. These comments also include gold-standard keywords. For instance, the notably higher distinctiveness of the gold-standard keywords in Article

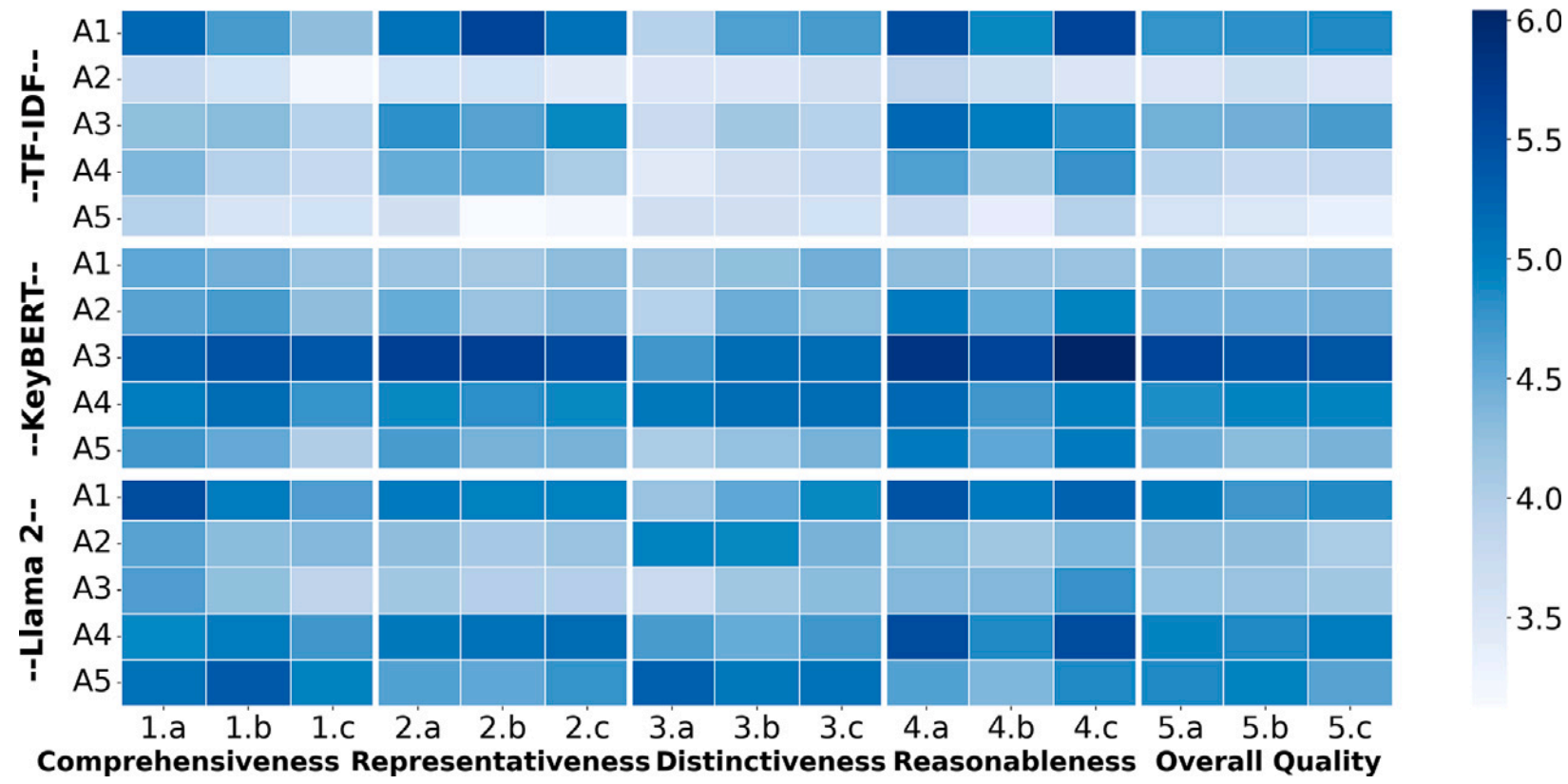

**Fig. 7.** Averaged evaluation scores of task instances from the preliminary experiment: Rows represent task instances (base articles indexed by A1–A5). Columns represent different questions. Cell colours indicate the average score for each question per task instance, with darker shades reflecting higher scores.

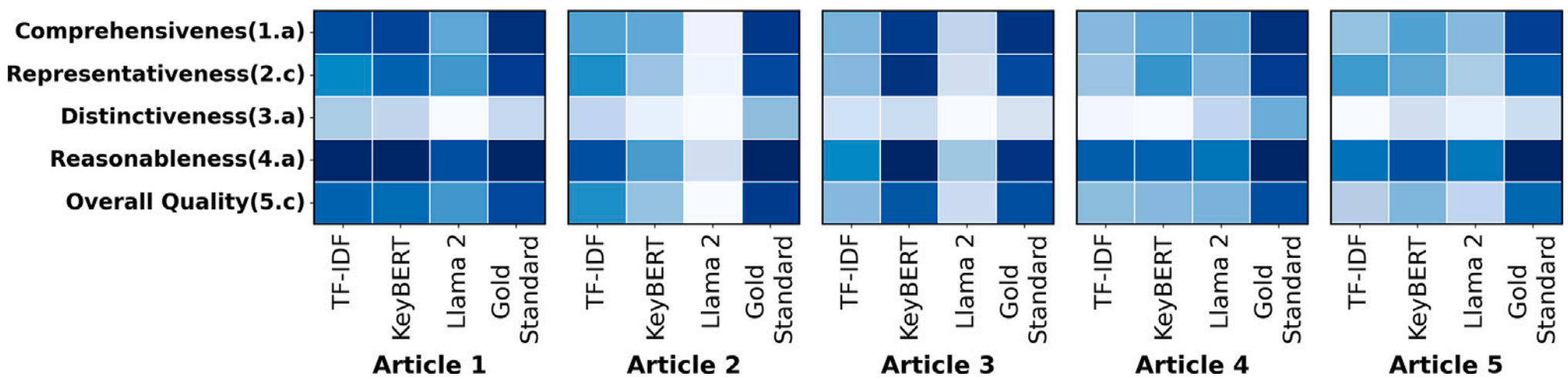

**Fig. 8.** Averaged evaluation scores of base articles from the main experiment: In each heat map, rows represent task questions, and columns represent keyword extraction methods. Cell colours indicate the average score for each question-method pair, with darker shades reflecting higher scores.

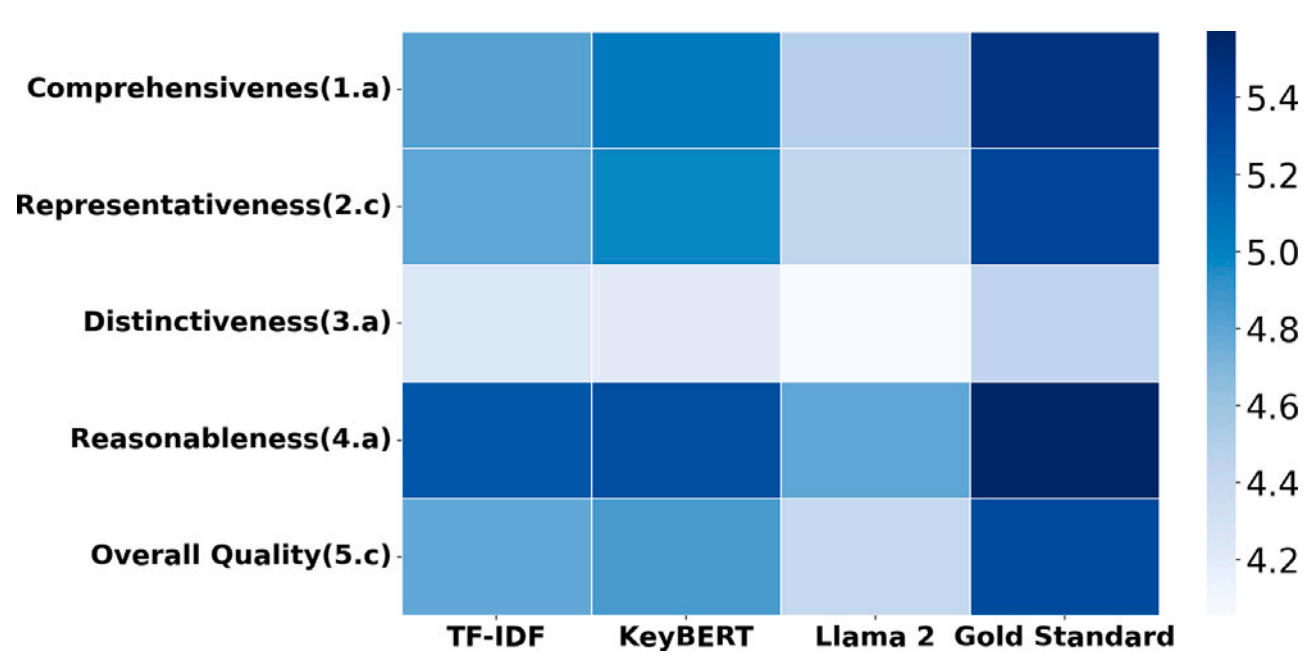

**Fig. 9.** Averaged evaluation scores across all base articles from the main experiment: Rows represent task questions, and columns represent keyword extraction methods. Cell colours indicate the average score for each question-method pair, with darker shades reflecting higher scores.

4 (see Fig. 8) may be attributed to its selection of all specific brand and celebrity names (see Table A.2 in Appendix A), which makes it less general compared to the algorithmic keyword sets. This suggests that despite the efforts to formulate clear task questions, the concept of distinctiveness may have confused participants and introduced variability into their ratings. Similar challenges are also noted in [29], which partially inspired our inclusion of *distinctiveness* as one of the properties of high-quality keywords. Due to the difficulty of defining evaluation standards, they use accuracy as an instrument to assess how distinctively the extracted keywords characterise the source document so that it can be used for more accurate document classification. Clarifying the measurement of distinctiveness could be a valuable area for future research.

The evaluation scores from the main experiment (Figs. 8 and 9) exhibit clear patterns in participants' assessments of the keyword properties, which repeat across different base articles and keyword sets. Specifically, *distinctiveness* consistently receives the lowest scores and *comprehensiveness* and *representativeness* are rated similarly. In addition to the above discussion about *distinctiveness*, these results reflect that the keyword sets perform better on *comprehensiveness* and *representativeness* than on *distinctiveness* in terms of participant evaluations. However, this difference might also be attributed to the limited number of keywords, or unclear definitions of the properties, and there is no explicit evidence of trade-offs among the properties based on the evaluation ratings or the comments. For instance, in Fig. 8, while participants rate Article 1 higher on *distinctiveness* compared to Article 5 on average, the ratings on other properties also increase, suggesting a general trend rather than distinct trade-offs. Further analysis using the Pearson correlation coefficients of the main experiment's results reinforces these findings. The analysis shows a stronger positive correlation among *comprehensiveness*, *representativeness* and *resonableness* than the correlation between the *distinctiveness* with other properties across all keyword sets. For example, *comprehensiveness* and *representativeness* are found to be moderately positively correlated in gold standard ratings, $r(N = 255) = .77, p < .01$, whereas the correlation between *comprehensiveness* and *distinctiveness* is less positive, $r(N = 255) = .38, p < .01$. Across different base articles, these correlations remain positive but generally decrease. For example, the correlation between *comprehensiveness* and *distinctiveness* drops to $r(N = 208) = .19, p < .01$ in ratings of Article 3.

### *4.3. Automatically versus manually selected keywords: the gap is still significant*

In the main experiment, participants occasionally rate an algorithmic keyword set higher than the gold-standard keywords. However, in the supplementary experiment, where participants rank all four keyword sets directly, the gold-standard keywords almost always come out on top (Fig. 10). Related to RQ1 and RQ3, this is further evidence of the importance of human annotations in algorithm evaluations, as also argued by [30]. In the case of real-world online recommendation scenarios, it highlights the value of including manually entered

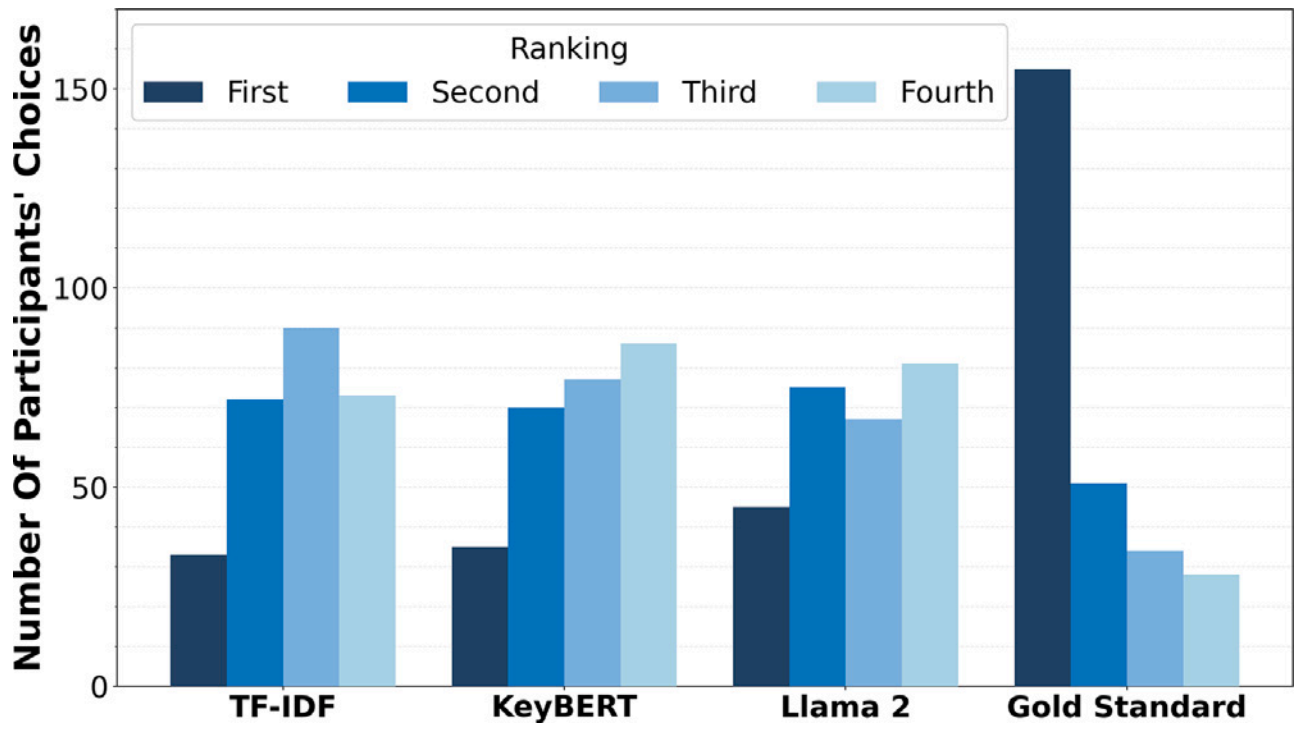

**Fig. 10.** Participant rankings of the TF-IDF, KeyBERT, Llama 2 and gold-standard keyword sets: The $x$-axis represents each extraction method, and the $y$-axis shows the frequency with which each keyword set is chosen for a specific rank.

keywords in the targeting, reinforcing the need for a human-in-the-loop approach to AI-driven strategies. A complementary perspective is provided by Fig. 11, which illustrates participants' choices of the best and the worst keyword sets regarding evaluation properties. There is a marked preference for gold-standard keywords, where the total number of participants selecting them as the best set is typically 1.5 to 4.6 times higher than algorithmic keywords. However, the differences among the algorithmic sets are not statistically significant, $\chi^2(df = 14, N = 1493) = 8.53$, $p = .86$.

### 4.4. The influence of algorithmic complexity on user preferences on keywords

Turning to RQ2, as shown in Fig. 7, participant preferences on the extracted keywords vary across base articles. For instance, for Article 1, it is evident that participants rate KeyBERT keywords (row 6) lower than TF-IDF (row 1) and Llama 2 (row 11). However, this pattern is completely reversed in the case of Article 3 (rows 3, 8, and 13). Despite these variations, a slight overall preference for KeyBERT emerges across all five base articles, with average ratings 14.8% higher than TF-IDF and 1.3% higher than Llama 2. This aligns with previous studies highlighting KeyBERT's ability to produce high-quality keywords [38,40]. However, the finding that it performs at least as well as Llama 2 in terms of direct participant ratings is novel. The same analysis is applied to the main experiment results, where each target property is evaluated using a single question. Fig. 8 confirms that the participant preferences vary across base articles. Notably, KeyBERT receives particularly high ratings for Article 3, which are even comparable to those of the gold standard keywords, receiving a 1.2% higher average score. In addition, KeyBERT slightly outperforms the gold standard over the properties of *representativeness*, *distinctiveness* and *reasonableness*. Fig. 9 presents the average scores for each extraction algorithm across all base articles. The trend is more pronounced than in the preliminary experiment: KeyBERT consistently outperforms the others, scoring on average 5.8% higher than TF-IDF and Llama 2, with the latter receiving the lowest ratings. When participants compare all keyword sets simultaneously, as shown in Figs. 10 and 11, slightly more participants select keywords extracted by Llama 2 over those produced by KeyBERT. However, although minor variations exist in the number of participants selecting the best keyword sets for each algorithm, as reported in Section 4.3, the differences are not statistically significant. These results indicate that a relatively simple but specialised method can outperform more advanced models in terms of user preferences when extracting keywords, especially for structured, high-quality corpora where contextual relevance is key [40]. As Llama 2 was prompted without explicit instructions regarding the desired properties, it likely relied on its general language understanding to identify the ten keywords that best describe each base article. A promising direction for future work would be to explore alternative prompting strategies that incorporate more specific and goal-oriented instructions to guide keyword extraction.

### 4.5. The influence of algorithmic complexity on user preferences on keyword-based recommendations

As shown in Fig. 12, participant ratings of article pairs recommended based on keywords extracted by TF-IDF, KeyBERT and Llama 2 are consistent with our findings from the preliminary, main and supplementary experiments. However, when compared with the state-of-the-art LLMs, the relative advantages and efficiency of the simpler methods are difficult to maintain across all base articles.

Questions 1 and 2 assess participants' interest in the base articles and their corresponding matched articles. As expected, participants' interest in the base articles (Q1) remains consistent across methods, since they are identical across all methods. In contrast, for the matched articles (Q2), those generated by Llama 2 receive the lowest ratings, while those generated by DeepSeek are rated the highest, followed by Mistral, KeyBERT, TF-IDF and Llama 3. Regarding the perceived relevance of article pairs (Q3-Q10, with Q7 reverse-coded), the average ratings for Llama 2 article pairs are consistent as the lowest across all relevance questions. TF-IDF achieves similar ratings as Llama 2.

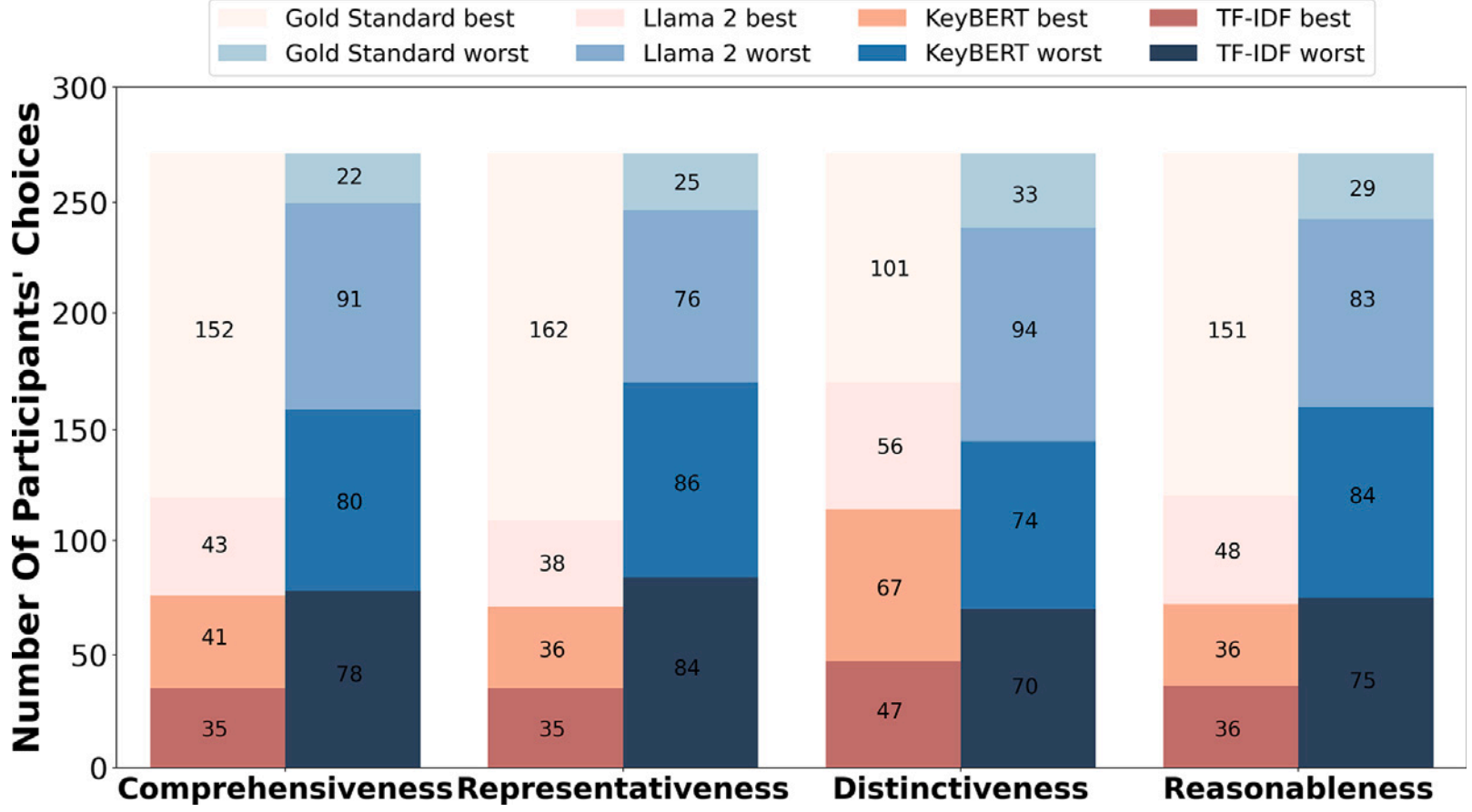

**Fig. 11.** Participant selections for the best and worst keyword sets across different evaluation properties: Orange bars represent the total number of participants' responses that selected the specific keyword set as the best set, while blue bars show selections for the worst set. The annotation numbers within the bars show the actual number of selections. (For interpretation of the references to colour in this figure legend, the reader is referred to the web version of this article.)

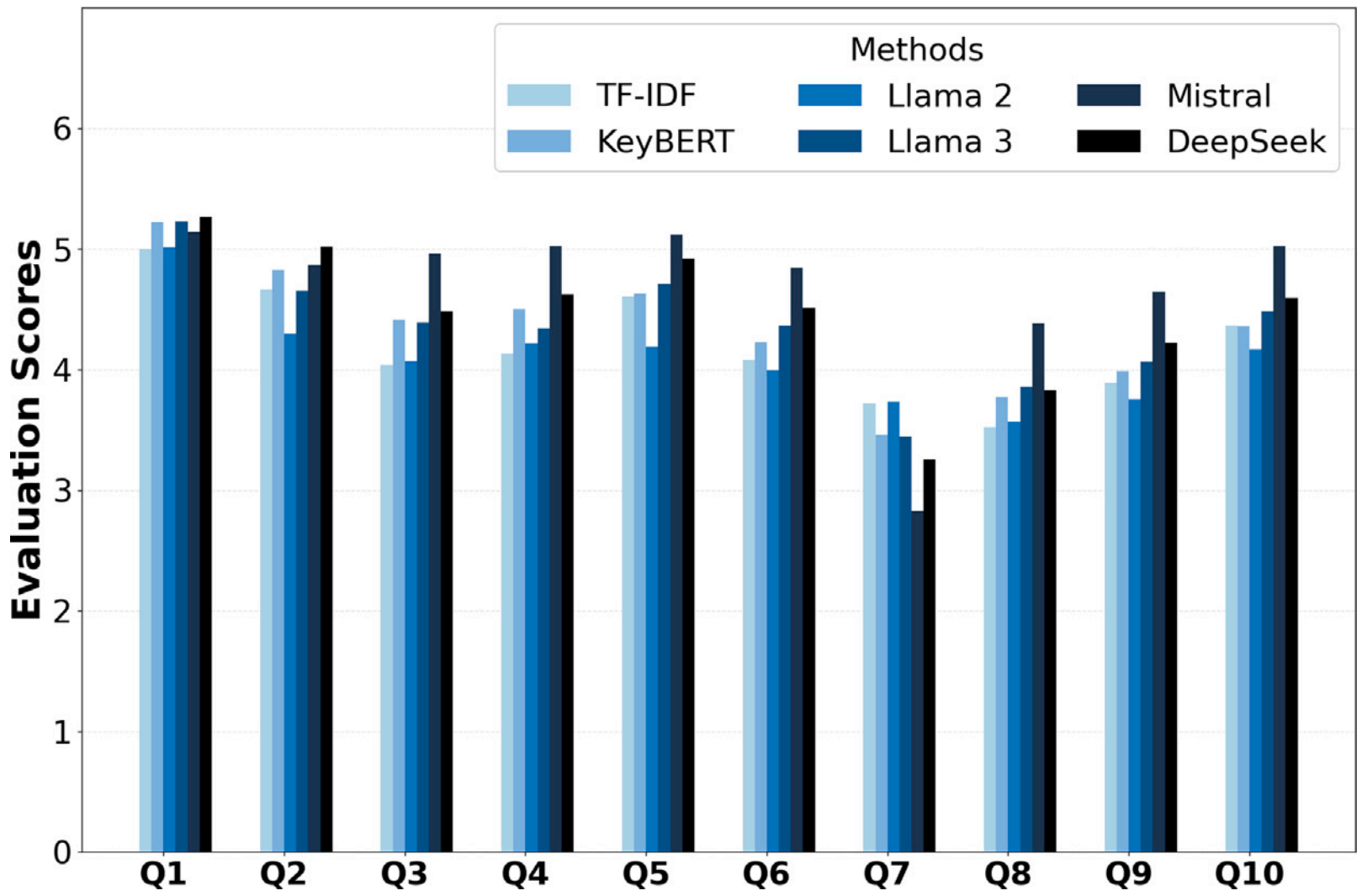

**Fig. 12.** Averaged evaluation scores of task instances from the document-recommendation experiment: The $x$-axis represents survey questions (see below and Appendix D) and the $y$-axis shows the average ratings across all task instances.

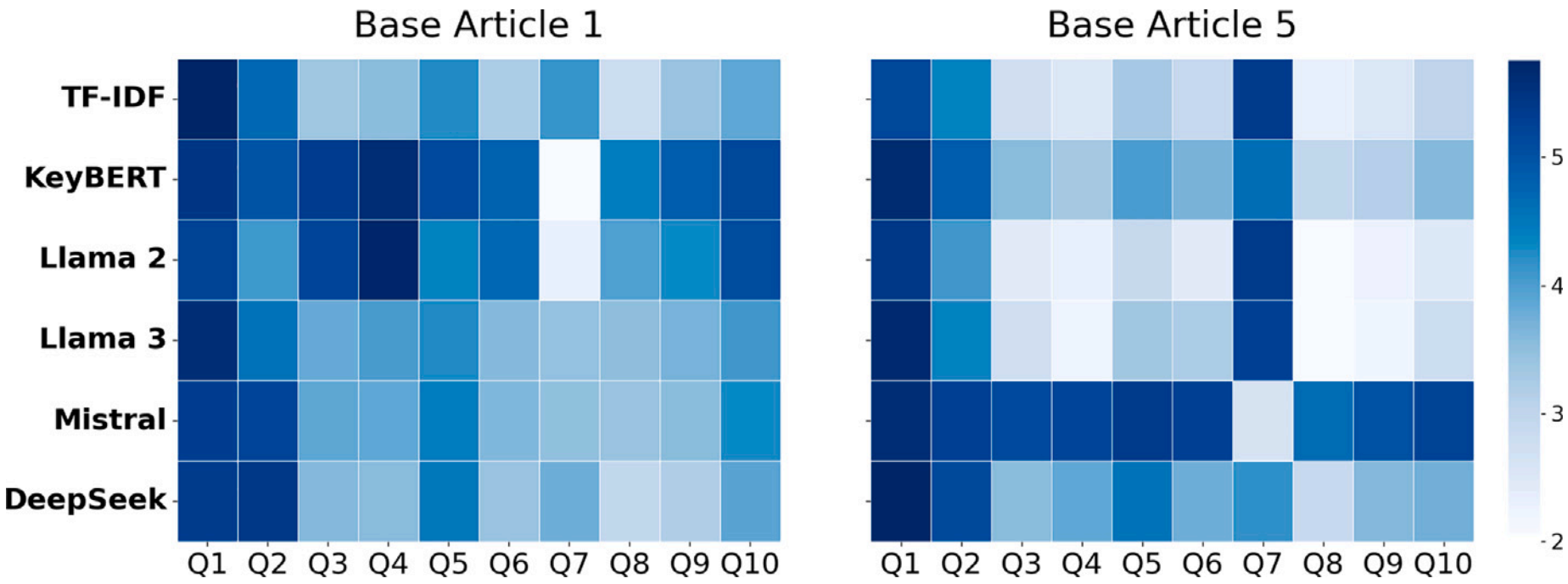

**Fig. 13.** Averaged evaluation scores for Base Articles 1 and 5 in the document-recommendation experiment: In each heat map, rows represent task questions, and columns represent keyword extraction methods. Cell colours indicate the average score for each question-method pair, with darker shades reflecting higher scores (reversed for Q7).

Except from Mistral, DeepSeek outperforms the other methods, but its advantage over KeyBERT is not statistically significant, even though it is the largest method among the six compared (e.g., for Q3, Q4, and Q5, $p$ = .85, .87 and .85, respectively). In addition, participants rate the article pairs generated by KeyBERT higher than those produced by Llama 3 in terms of relevance (Q3, $p$ = .05) and topical similarity (Q4, $p$ = .05), highlighting the efficiency of KeyBERT in these aspects. Notably, Mistral achieves significantly higher participant ratings across all relevance-related questions. For instance, compared to DeepSeek, Mistral obtains substantially higher ratings on Questions 8, 9, and 10 ($p < .05$, $p = .05$ and $p < .05$).

As discussed in Sections 4.2 and 4.4, participant preferences vary considerably across articles. This finding is also supported by the results of the downstream experiment. Fig. 13 provides an example of the comparisons of participant ratings across different base articles. For Article 1, KeyBERT and Llama 2 receive significantly higher ratings than other methods. For example, compared with Mistral, the average ratings for KeyBERT on Questions 3 and 4 are 31.5% ($p < .01$) and 38.2% ($p < .01$) higher, and for Llama 2, 28% ($p < .05$) and 40.7% ($p < .001$) higher. However, this pattern is completely reversed for Article 5, where Mistral receives significantly higher ratings than all other models. Participants' comments on the matched articles for Article 5 indicate that models such as KeyBERT and DeepSeek tend to overemphasise location-related keywords, whereas Mistral successfully identifies keywords and matches based on broader and more holistic topical contexts. These findings highlight how participants' evaluations are shaped by different articles and reinforce the need for human-in-the-loop assessment frameworks over end-to-end automated benchmarking evaluations.

**Table 1**
Comparisons of keywords generated by TF-IDF, KeyBERT, Llama 2, Llama 3, Mistral, and DeepSeek against gold-standard keywords. Metric scores for each method are averaged across five base articles.

| Keywords | Precision | Recall | Cosine Similarity | Edit Distance |
|---|---|---|---|---|
| TF-IDF | 0.56 | 0.56 | 0.89 | 3.10 |
| KeyBERT | 0.54 | 0.54 | 0.90 | 2.92 |
| Llama 2 | 0.40 | 0.40 | 0.90 | 2.90 |
| Llama 3 | 0.60 | 0.58 | 0.90 | 2.78 |
| Mistral | 0.60 | 0.58 | 0.92 | 2.82 |
| DeepSeek | 0.64 | 0.64 | 0.94 | 3.26 |

### 4.6. Precision-based benchmarking and human perceptions

To quantitatively assess the algorithmic keyword sets, they are compared against the gold-standard keywords using standard quality metrics (see Table 1), addressing RQ2. The first two columns show the *precision* and *recall*, which result in identical figures in this case because both algorithmic and gold-standard sets have a fixed length of ten. Higher precision and recall indicate stronger overlap with human-annotated keywords, reflecting the effectiveness and comprehensiveness of the evaluated algorithm in extracting keywords that closely resemble human choices. However, this basic method fails to capture semantic similarity. For instance, it treats the inclusion of 'panda' and 'deterioration' as equally incorrect substitutions for the gold-standard word 'bear'. To address this, a Word2Vec [78] model trained on all news articles from the candidate pool is used to compute the *cosine distance* between each algorithmic and the gold-standard keyword set. Correspondingly, a higher cosine distance reflects greater semantic alignments between the compared keyword sets. Finally, the *edit distance* concerning position replacements is calculated, to account for the order of the keywords. Algorithmic keyword orders are based on their weights during automatic extraction, whereas the order of gold-standard keywords reflects the frequency of participant selections. Therefore, if the algorithms assign lower weights to the keywords that are most frequently chosen by participants, they still deviate from the gold standard, even though the keyword is extracted accurately. Accordingly, greater edit distances indicate less similarity in the content and order of the keywords.

The quantitative analyses combining precision, recall, cosine distance, and edit distance for TF-IDF, KeyBERT and Llama 2 align with the findings from the qualitative user experiments. When evaluated against the gold-standard keywords, KeyBERT-generated keywords are found to be 'closer' to those selected by participants, which may explain why they have higher ratings. KeyBERT's ability to capture semantics and context has been highlighted in previous work [38,40]. Regarding research question RQ3, this suggests that KeyBERT is more likely to resonate with target audiences and align with the contextual nuances of advertising campaigns, thereby enhancing advertising efficiency and return on investment [21,22]. However, a clear misalignment emerges between algorithmic benchmarking and participant perceptions in the downstream experiments. According to these precision-based benchmarking metrics, recent state-of-the-art LLMs demonstrate stronger capabilities in extracting keywords that are closer to gold-standard references and in capturing contextual information. In particular, keywords extracted by DeepSeek achieve the highest scores on precision, recall, and cosine similarity. Nevertheless, as discussed in Section 4.5, there is no significant evidence that participants generally prefer the matched articles based on these more 'precise' keywords extracted by DeepSeek. This misalignment highlights the gap between automated benchmarking and real user perception and experience in practice.

### 4.7. Implications for real-world applications

As discussed in Section 2, real-world applications such as online advertising auctions occur at a rate of hundreds of thousands per second, requiring each auction to have a minimal computational footprint. Moreover, auctions must be completed rapidly to avoid disrupting the user experience. Consequently, in relation to RQ3, this implies that the algorithms used in practical applications should be lightweight in terms of both execution time and computational cost, or otherwise rely on GPUs or cloud-based infrastructures, which requires substantial financial and technical resources for deployment and maintenance. Consequently, the largest LLMs may not be a practical alternative, due to their high inference costs and relatively slower execution speeds. Fig. 14 presents the processing times for TF-IDF, KeyBERT and Llama 2 in extracting keywords from the five task articles over 100 epochs. To eliminate the influence of caching and I/O operations, each algorithm was prepared by five warm-up runs to ensure that the cache and compiler are fully initialised. In addition, all runtime measurements exclude document reading and result writing, recording only the core processing time. For Llama 2, apart from the extraction prompt, no explicit instruction is given to discard previous conversation histories; instead, the model is restarted for every keyword extraction to ensure a clean context. However, as all three algorithms were executed on a local device using a CPU, the processing times may be slower compared to those with GPUs or cloud-based services. Although employing GPUs can reduce the computational time of LLMs, similar acceleration benefits would also apply to the simpler methods. Therefore, even if the efficiency gap narrows in such cases, the simpler approaches would still remain computationally lighter. In addition, variations in GPU configurations inevitably lead to differences in computational cost results. For this reason, we use CPU-based execution as a fundamental and reproducible baseline to illustrate the efficiency comparison in a controlled and consistent manner. As shown in the figure, TF-IDF is the least computationally intensive, followed by KeyBERT, whereas Llama 2 is significantly more time-consuming, operating on a completely different scale from the other two.

Reflecting on the evaluated properties, comprehensiveness and reasonableness are the most crucial for keyword-based applications. They ensure that keywords capture various aspects of the source document, offering a solid understanding of its semantics. While distinctiveness and representativeness contribute to keyword uniqueness, they are less relevant for applications like contextual advertising, where the primary focus is on document-ad matching. However, if these two properties are excluded, KeyBERT becomes even more compelling, as its performance is closest to that of the alternatives when considering distinctiveness. In both the quantitative and qualitative evaluations, KeyBERT ranks at least as high as Llama 2 in most cases compared to the gold standard and participant evaluations, with approximately 0.3% of the computational resources needed by Llama 2. Additionally, KeyBERT retains TF-IDF's advantage of being relatively easy to implement and deploy, making it an accessible, efficient keyword extraction method for advertisers and marketers, as well as a valuable tool for various other real-world applications.

Related to RQ1, assessing alternative phrasings for evaluation properties in the preliminary experiment ensures clear survey questions, minimising the influence of experimental variables, and enhances the authenticity and independence of participant responses. This approach, which strengthens user feedback on keyword extraction methods, provides valuable insights for user-centred studies, where surveys are a key tool for guiding them in interactions and collecting user experiences.

Lastly, considerable trade-offs exist between algorithmic performance and user preferences, due to the influence of several factors. First, the predicted economic value of a user clicking an ad varies by product categories. Second, the supply of relevant media contexts differs, with articles about Premier League football being more abundant than those on fly fishing [79]. Third, the degree of congruence needed for the desired advertising effect also matters: an advertiser targeting 'sugar-free ice cream' may not want articles on food and drink in general. Therefore, regardless of the algorithm, it is advantageous if it can be tuned for both greater accuracy, higher user preference and faster computation. With KeyBERT, this can be achieved by adjusting the vocabulary size, or the embedding size. LLMs can support similar trade-offs, such as through model distillation where a simpler model is derived from a more complex one while retaining much of its performance on specific tasks [80]. However, this requires more advanced technical resources.

## 5. Conclusion and future directions

Keyword extraction continues to be a central task in natural language processing. While recent advancements in language modelling contribute to groundbreaking progress across various domains, they come at the cost of increased computational complexity and higher hardware requirements. Although neural models are often fine-tuned with human feedback, less effort has been placed on analysing the user-perceived performance of downstream algorithms that depend on the algorithms or LLM's output. For keyword extraction, simpler models can still be competitive to some LLMs, particularly when integrating human perceptions in evaluation for user-centric applications such as

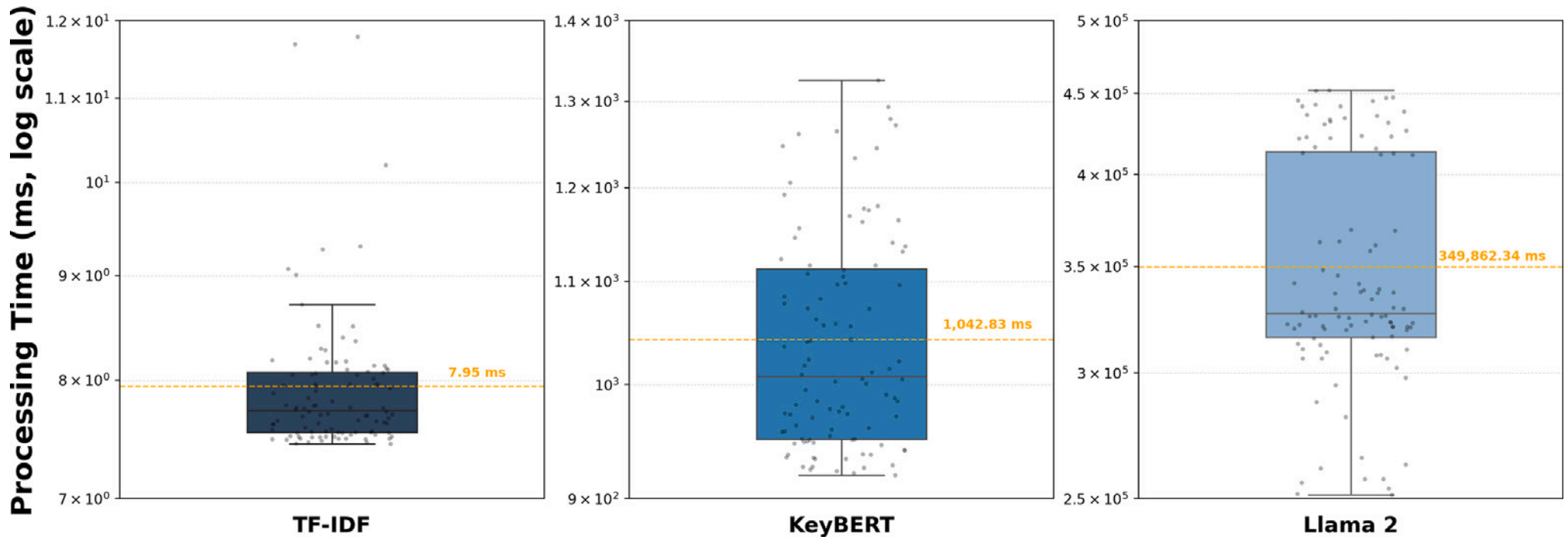

**Fig. 14.** Processing time of extracting keywords over 100 epochs, reported in milliseconds: Each grey dot corresponds to the time spent for extracting keywords for all five base articles per epoch. Horizontal orange lines show the average processing time across these 100 epochs. Box plots show the distribution of these processing times using the same visualisation style as in Figs. 3 and 5. (For interpretation of the references to colour in this figure legend, the reader is referred to the web version of this article.)

contextual advertising ot information retrieval tasks that require the efficient processing of large amounts of text.

Beyond typical precision-based evaluation metrics, this study emphasises user perceptions in assessing keyword extraction methods represented by TF-IDF, KeyBERT, Llama 2, Llama 3, Mistral and DeepSeek, and refines reliable assessment tools through a combination of qualitative and quantitative human-in-the-loop methods. To address RQ1, this study demonstrates the value of analysing user perception for such evaluations in real-world applications, using non-experts to reduce resources and time, while tackling subjective challenges through a survey-based approach. Evaluation properties are developed and implemented to ensure consistent feedback and minimise unexpected shifts in user responses. Two novel analytical approaches, horizontal and vertical analysis, are introduced for analysing question phrasing consistency, offering valuable insights for improving experimental design. In addition, the preliminary experiment validates the feasibility of the evaluation process, serving as a foundation for subsequent experiments and improves our understanding of user perceptions.

To address RQ2, the qualitative experiments provide valuable insights into the performance of the compared extraction methods and reveal variations in participant preferences across different base articles. Overall, KeyBERT demonstrates greater efficiency over TF-IDF and Llama 2 in both the keyword and downstream experiments, whereas Mistral significantly outperforms all other models in the document-recommendation task. Moreover, KeyBERT slightly outperforms the gold standard in terms of the assessed properties, including *reasonableness*. Both quantitative and qualitative results highlight the efficiency of KeyBERT in producing high-quality keywords favoured by participants. The results also indicate some instances where algorithmic keyword sets are slightly preferred over the gold-standard keywords. However, the overall findings show a clear user preference for the gold standard across all evaluation properties. In contrast, differences between algorithmic sets are not statistically significant, underscoring the enduring reliability of manually extracted keywords.

Finally, in terms of RQ3, these findings provide insights for researchers and practitioners to reflect on the balance between performance optimisation and user-centred evaluation, particularly for real-world applications where computational efficiency is critical and the primary goal is to meet end-user needs. The rapid pace and large scale of digital advertising demand fast and accurate algorithms, and computational costs are particularly crucial when evaluating keyword extraction algorithms for contextual advertising. In this study, KeyBERT has proven to be an attractive option, as it offers competitive performance compared to Llama 2 and, in some cases, other LLMs, while requiring significantly fewer computational resources. The ease of implementation and deployment further enhances its appeal to practitioners seeking efficient solutions. The effectiveness of this relatively lightweight model highlights the long-overlooked disconnect between the development of keyword extraction methods and their real-world applications: rather than solely pursuing incremental improvements in precision, it is worth considering what real users truly value in practice.

In conclusion, this study underscores the importance of human-in-the-loop evaluation in keyword extraction for contextual advertising and calls for more attention to user perceptions in assessing algorithmic performance. Through both precision-based and perception-based experimental analysis on these representative keyword extraction methods, including LLMs, simpler methods like KeyBERT show competitive performance with greater computational efficiency for real-world applications. Future studies could expand this work by exploring additional keyword extraction algorithms, such as unsupervised graph-based algorithms like TextRank [5], or topic-based approaches like LDA [81], as well as other LLMs. Finally, while this study focused on comparing algorithm-generated keywords by surveying users on pre-defined evaluation properties, future research could further explore how users perceive these properties. This may include asking participants to identify which properties they consider most important and to comment explicitly on perceived relationships among them. It should be noted, however, that such perceptions are often multi-dimensional and intertwined, and may even contain contradictions — as illustrated by participants giving lower ratings on *distinctiveness* even to gold-standard keywords selected by humans.

## 6. Limitations

Participants were recruited based on minimal screening criteria to mitigate potential bias. However, the participant pool provided by the Prolific platform may have limitations in ensuring diversity and representing the general population among the participants. Moreover, as noted in the participant feedback, some individuals speculated about the purpose of the experiments, potentially influencing their evaluations. This behavioural shift may be partly explained by the Hawthorne effect, referring to individuals' tendency to alter their behaviour when they are aware of being observed.[9] Although the practical implications of this phenomenon remain uncertain [83], it is clear that such effects cannot be fully eliminated.

When creating the gold-standard keyword sets from the manually annotated keywords, those most frequently chosen by participants are prioritised. This can result in semantically similar words being included in the set, leaving less room for additional words that may improve

[9] The Hawthorn effect was first documented in a series of experiments conducted between 1924 and 1933 at a factory in Hawthorne, Illinois, which found that workers tended to improve their productivity when they knew they were being studied [82].

**Table A.1**
URLs of the five base articles.

| Id | URL |
|---|---|
| 1 | https://www.theguardian.com/society/2022/may/25/more-than-one-in-10-young-women-now-identify-lesbian-gay-bisexual-or-other |
| 2 | https://www.theguardian.com/news/audio/2022/jul/29/euro-2022-and-the-future-of-womens-football |
| 3 | https://www.theguardian.com/world/2022/may/31/14th-century-samurai-sword-found-in-car-at-swiss-border |
| 4 | https://www.theguardian.com/music/2022/oct/26/kanye-west-escorted-out-skechers |
| 5 | https://www.theguardian.com/world/2022/jul/21/an-an-worlds-oldest-captive-male-giant-panda-dies-in-hong-kong-zoo-aged-35 |

**Table A.2**
The keywords generated by TF-IDF, KeyBERT, Llama 2 and participants for five base articles.

| Article | TF-IDF | KeyBERT | Llama 2 | Gold Standard |
|---|---|---|---|---|
| 1 | people, gay, bisexual, uk, heterosexual, identifying, lesbian, figure, identify, young | heterosexual, bisexual, lesbian, gay, gender, straight, sexual, percentage, ethnicity, footballer | lesbian, gay, bisexual, other, female, young, men, age, sexual, orientation | bisexual, lesbian, gay, uk, women, young, orientation, sexual, openness, heterosexual |
| 2 | football, people, uk, girl, game, woman, moore, muslim, organisation, passion | lioness, footballer, football, sport, england, woman, girl, playing, fa, hannah | success, girls, uk, guardian, suzanne, football, fa, trolling, spain, brighton | football, lionesses, women, uk, accessible, muslim, girls, role, model, media |
| 3 | custom, swiss, sword, authority, driver, franc, found, object, antique, investigation | katana, sword, custom, swiss, antique, smuggled, franc, fine, samurai, zurich | swiss, authorities, sword, japan, discovered, book, contract, invoice, driver, daughter | swiss, sword, customs, smuggled, fines, investigation, criminal, antique, samurai, authorities |
| 4 | ye, skechers, antisemitic, recent, comment, company, adidas, gap, longer, longtime | kanye, adidas, rapper, supremacist, footwear, skechers, fashion, brand, white, gap | skechers, kanye, west, unannounced, offices, artist, fashion, statement, brand, antisemitic | skechers, antisemitic, kanye, west, adidas, ye, gap, supremacists, unauthorised, conspiracy |
| 5 | park, ocean, kong, chinese, year, giant, hong, panda, friendship, died | panda, jia, chinese, zoo, kong, hong, sichuan, oldest, ocean, ying | panda, oldest, hong, an, died, thursday, zoo, china, jia, ying | panda, kong, hong, oldest, died, zoo, an, euthanised, ocean, park |

*representativeness* and *distinctiveness*. However, 25.8% of the participant-selected keyword sets contain fewer than ten words. This may partly be due to their fatigue or boredom, but it may also indicate the relative importance of the remaining words, even if they carry some semantic overlap. Conversely, 6.8% of the sets exceeding ten words are adjusted by removing duplicates in the final frequency calculation. This procedure may introduce bias into the results, as it is difficult to balance frequency standards when, for example, one participant selects five words with three duplicates, while another includes the same word three times within a 20-word list. Despite these limitations, the gold-standard keyword sets consistently outperform the algorithmic keyword sets, and further refinement of their construction would likely accentuate this performance gap.

As described in Section 3, standard implementations of algorithms and models were applied. While domain-specific data could have improved performance, it would also have increased the number of experimental variables and introduced additional bias. Future work should therefore explore how domain adaptation, fine-tuning strategies, and larger training corpora influence performance.

To minimise potential task fatigue and maintain focus on our primary research objectives, the keyword experiments (preliminary, main and supplementary experiments) were restricted to three mainstream algorithms and all experiement focused exclusively on 1-gram keywords. These choices may limit the generalisability and scope of the findings, although this impact is partly mitigated by allowing participants to evaluate keywords as collective sets.

Algorithmic development, particularly in the Large Language Models field, is advancing rapidly, making it inevitable that newer and more powerful models will emerge, as occurred during the course of this study. To address this, we conducted an additional document-recommendation experiment simulating real-world contextual tasks, using an expanded set of LLMs, including Mistral, Llama 3 and DeepSeek. Extending this setup with upcoming models could strengthen and validate these findings even further.

## CRediT authorship contribution statement

**Jingwen Cai:** Writing – review & editing, Writing – original draft, Visualization, Validation, Software, Resources, Methodology, Formal analysis, Data curation, Conceptualization. **Sara Leckner:** Writing – review & editing, Writing – original draft, Validation, Methodology, Funding acquisition, Formal analysis, Conceptualization. **Johanna Björklund:** Writing – review & editing, Writing – original draft, Validation, Resources, Methodology, Funding acquisition, Formal analysis, Conceptualization.

## Ethical statement

The experiments involve the recruitment of participants via Prolific and the collection of demographic information. This process strictly adheres to Prolific's privacy policies and ensures the anonymity of participants. Demographic data collection involves only basic information, including gender (including non-binary options), age (divided into age intervals), and level of education. These demographic data are used for research purposes only and are only accessible to the researchers involved in the study.

## Declaration of Generative AI and AI-assisted technologies in the writing process

During the preparation of this paper, the authors used DeepL and ChatGPT in order to translate the authors' texts in their native languages into English, and copy-edit texts for grammar and clarity. After using these tools/services, the authors reviewed and edited the content as needed and take full responsibility for the publication.

## Declaration of competing interest

The authors declare that they have no known competing financial interests or personal relationships that could have appeared to influence the work reported in this paper.

## Appendix A. Base articles and keyword lists

The URLs of the five base articles and their keywords generated by TF-IDF, KeyBERT and Llama 2 are listed in Tables A.1 and A.2, along with the gold-standard keywords selected by the participants from the preliminary experiment for each article.

## Appendix B. Participant demographics

The tables below list the distribution of the demographic information collected from the participants who took part in the preliminary, main, supplementary and downstream experiments (see Tables B.1–B.3).

## Appendix C. Controlling for task fatigue

*In your experience with answering questions about these news articles, please share your **fatigue level*** (See Table C.1).

(a) Exhausted — I could only focus on one article.
(b) Somewhat fatigued — I started losing focus after the second article.
(c) Just right — Three articles is a manageable amount for me.

**Table B.1**
Total number of participants and distribution of gender, respectively.

| Experiment | | | | Participants |
|---|---|---|---|---|
| Preliminary | | | | 196 |
| Main | | | | 264 |
| Supplementary | | | | 92 |
| Downstream | | | | 303 |
| **Gender** | **Preliminary experiment** | **Main experiment** | **Supplementary experiment** | **Downstream experiment** |
| Woman | 67.9% | 61.0% | 63.0% | 53.5% |
| Man | 29.1% | 35.6% | 33.7% | 46.2% |
| Nonbinary | 2.5% | 2.3% | 1.1% | 0.3% |
| Prefer not to disclose | 0.5% | 0.4% | 0.0% | 0.0% |
| Prefer to self-describe | 0.0% | 0.7% | 2.2% | 0.0% |

**Table B.2**
Distribution of the educational level of the participants in experiments.

| Education level | Preliminary experiment | Main experiment | Supplementary experiment | Downstream experiment |
|---|---|---|---|---|
| Less than a high school diploma | 1.5% | 1.1% | 0.0% | 1.0% |
| High school diploma or equivalent | 20.4% | 20.8% | 25.0% | 25.4% |
| College degree | 26.0% | 23.5% | 21.7% | 28.0% |
| Graduate degree | 50.6% | 51.9% | 52.2% | 43.6% |
| Other | 0.5% | 1.9% | 0.0% | 2.0% |
| Prefer not to disclose | 1.0% | 0.8% | 1.1% | 0.0% |

**Table B.3**
Distribution of the age of the participants in the experiments.

| Age | Preliminary experiment | Main experiment | Supplementary experiment | Downstream experiment |
|---|---|---|---|---|
| <18 | 0.0% | 0.0% | 0.0% | 0.0% |
| 18–24 | 14.3% | 13.3% | 13.1% | 5.9% |
| 25–29 | 15.8% | 17.4% | 18.5% | 8.9% |
| 30–34 | 19.9% | 14.8% | 16.3% | 13.2% |
| 35–39 | 12.2% | 12.5% | 17.4% | 10.2% |
| 40–44 | 9.7% | 11.0% | 13.1% | 11.9% |
| 45–49 | 0.6% | 0.9% | 0.3% | 11.2% |
| 50–54 | 11.2% | 5.7% | 4.3% | 13.2% |
| 55–59 | 3.1% | 6.8% | 4.3% | 10.2% |
| 60–65 | 5.1% | 6.4% | 0.0% | 7.0% |
| >65 | 3.1% | 4.2% | 7.6% | 8.3% |
| Prefer not to disclose | 0.0% | 0.0% | 1.1% | 0.0% |

**Table C.1**
Distribution of the degree of fatigue of the participants in the preliminary experiments. 182 out of 196 participants provided feedback on their fatigue level.

| Fatigue level | Participants |
|---|---|
| Exhausted (1 article) | 3 |
| Somewhat fatigued (2 articles) | 37 |
| Just right (3 articles) | 98 |
| Not too bad (4 articles) | 32 |
| Energised (5 articles) | 7 |
| Extremely energised (6 or more articles) | 5 |

(d) Not too bad — I'm up for another one.
(e) Energised — I'm not tired at all and can handle up to five articles with full energy.
(f) Extremely energised — I can easily concentrate on even more articles.

## Appendix D. Survey questions

The following survey questions were used in the downstream experiment to measure participants' perspectives across three aspects: interest in the articles (Questions 1–2), perceived relevance and topical similarity between articles (Questions 3–10), and open-ended feedback (Question 11). Questions 1 to 10 were rated on a 7-point Likert scale ranging from 'strongly disagree' to 'strongly agree', while Question 11 was optional and allowed participants to provide free-text comments.

1. I find the first article interesting.
2. I find the second article interesting.
3. The two articles are related.
4. The two articles share similar topics.
5. If someone like me is interested in the first article, they would be interested in reading the second article as well.
6. The two articles appear to be specifically chosen for each other.
7. The two articles are unrelated.
8. The two articles together seem to cover the main aspects of the same topic.
9. The two articles seem to well represent the same area or topic.
10. It seems reasonable that these two articles are matched in terms of their contexts.
11. If you think these two articles should not be matched, please explain why.

## Appendix E. Task instances

Here we summarise the task instances and introduce the webpages used in our experiments (See Table E.1 and Figs. E.1–E.4.).

**Table E.1**
The number of task instances assigned to each participant in the experiments.

| Experiment | Number of task instances assigned to each participant | Number of task instances in the pool | Number of questions for each task instance |
|---|---|---|---|
| Preliminary | **3** | **30** (5 *base articles* × 3 *keyword sets* × 2 *question groups*) | 9 |
| Main | **4** | **20** (5 *base articles* × 4 *keyword sets*) | 5 |
| Supplementary | **3** | **5** (*all keyword sets showed on the same instance*) | 9 |
| Downstream | **5** | **150** (5 *base articles* × 5 *matched articles* × 6 *methods*) | 11 |

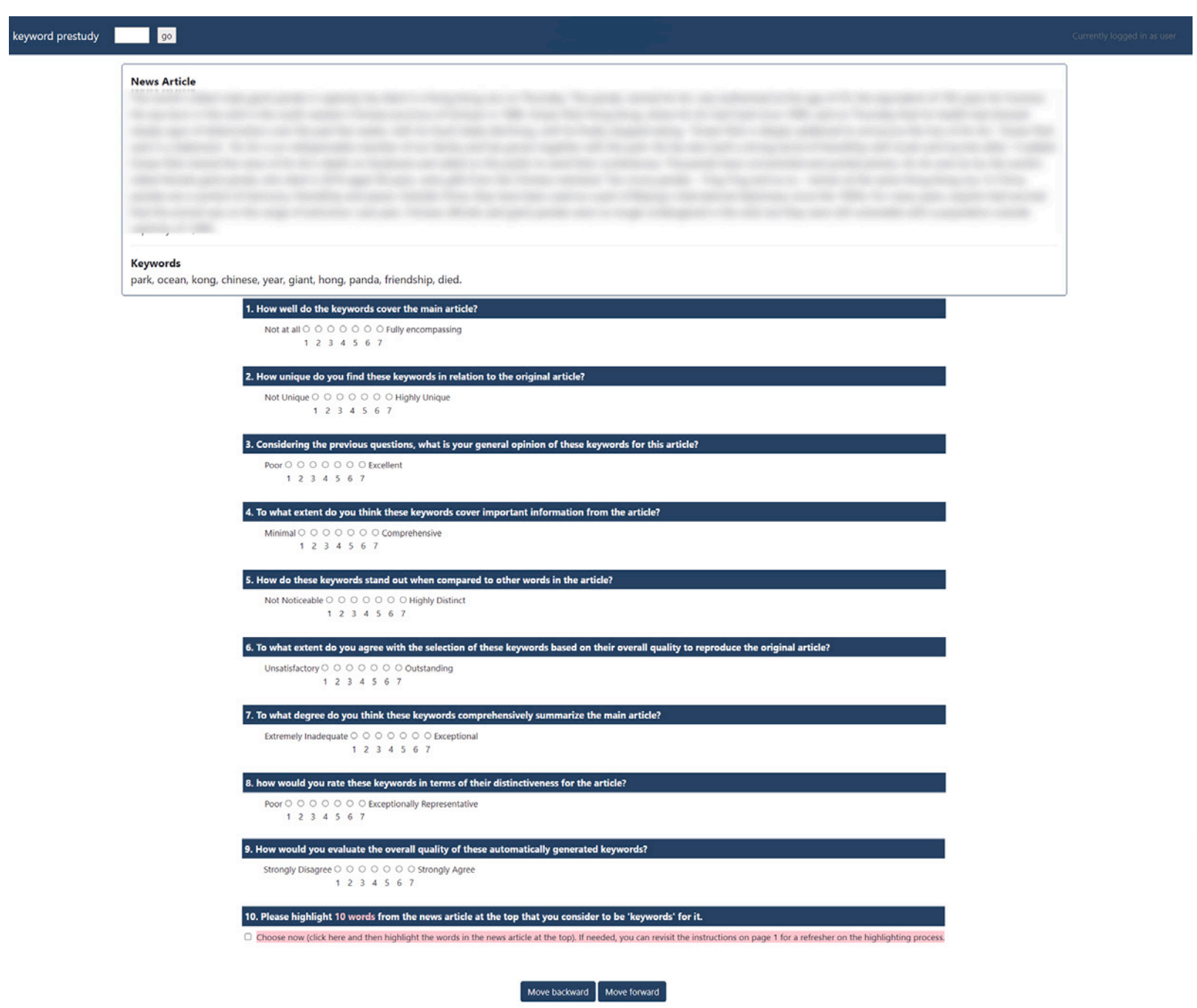

**Fig. E.1.** An example of the task instance page for the preliminary experiment: It contains a random combination of a base article and a set of 10 keywords sampled from the initial instance pool, nine 7-point Likert-scale questions covering one property group, and a highlighting task asking participants to choose 10 keywords they think are appropriate for the provided article.

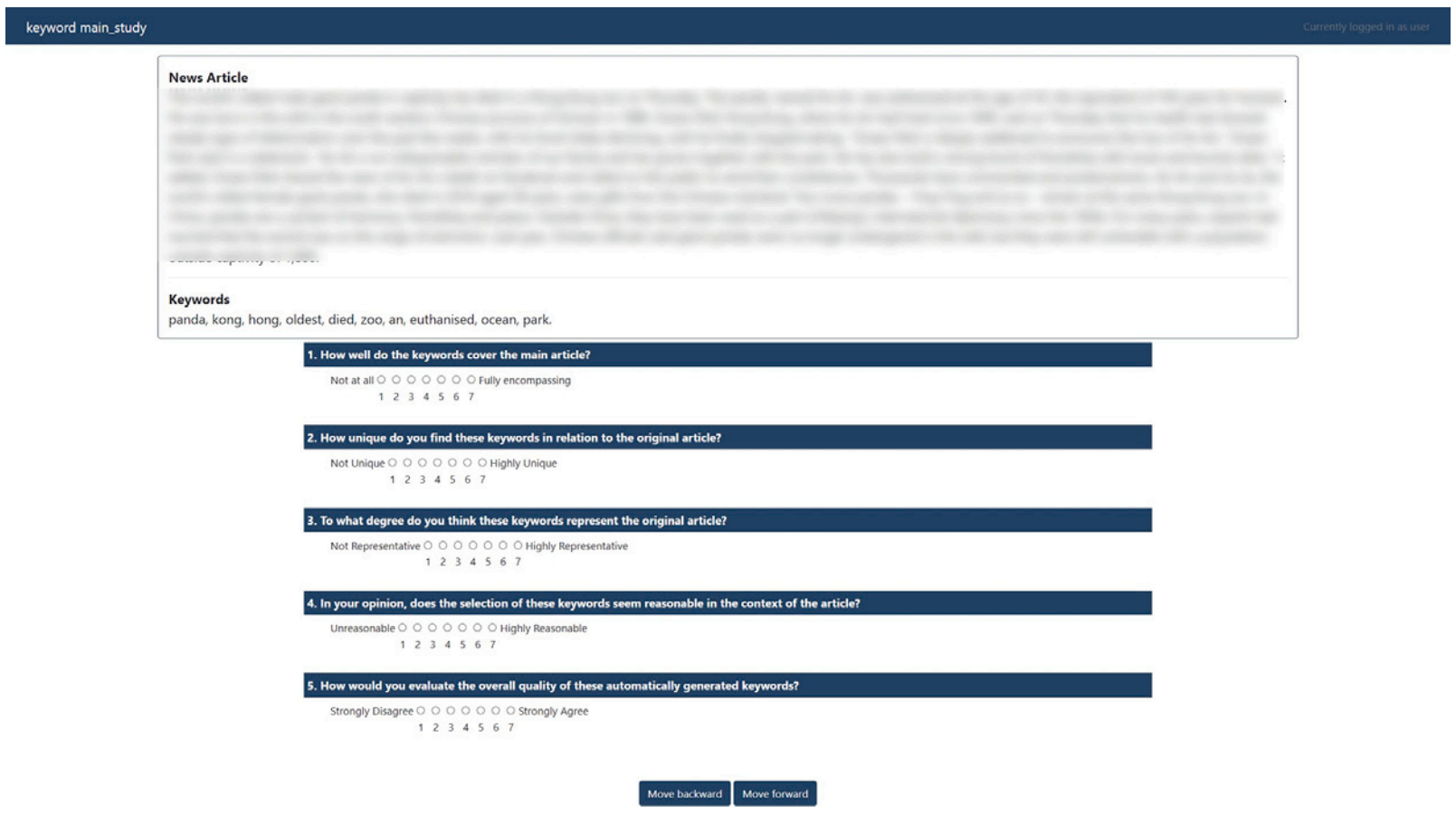

**Fig. E.2.** An example of the task instance page for the main experiment: It contains a random combination of a base article and a set of 10 keywords sampled from the task instance pool (including gold-standard keywords), and five 7-point Likert-scale questions.

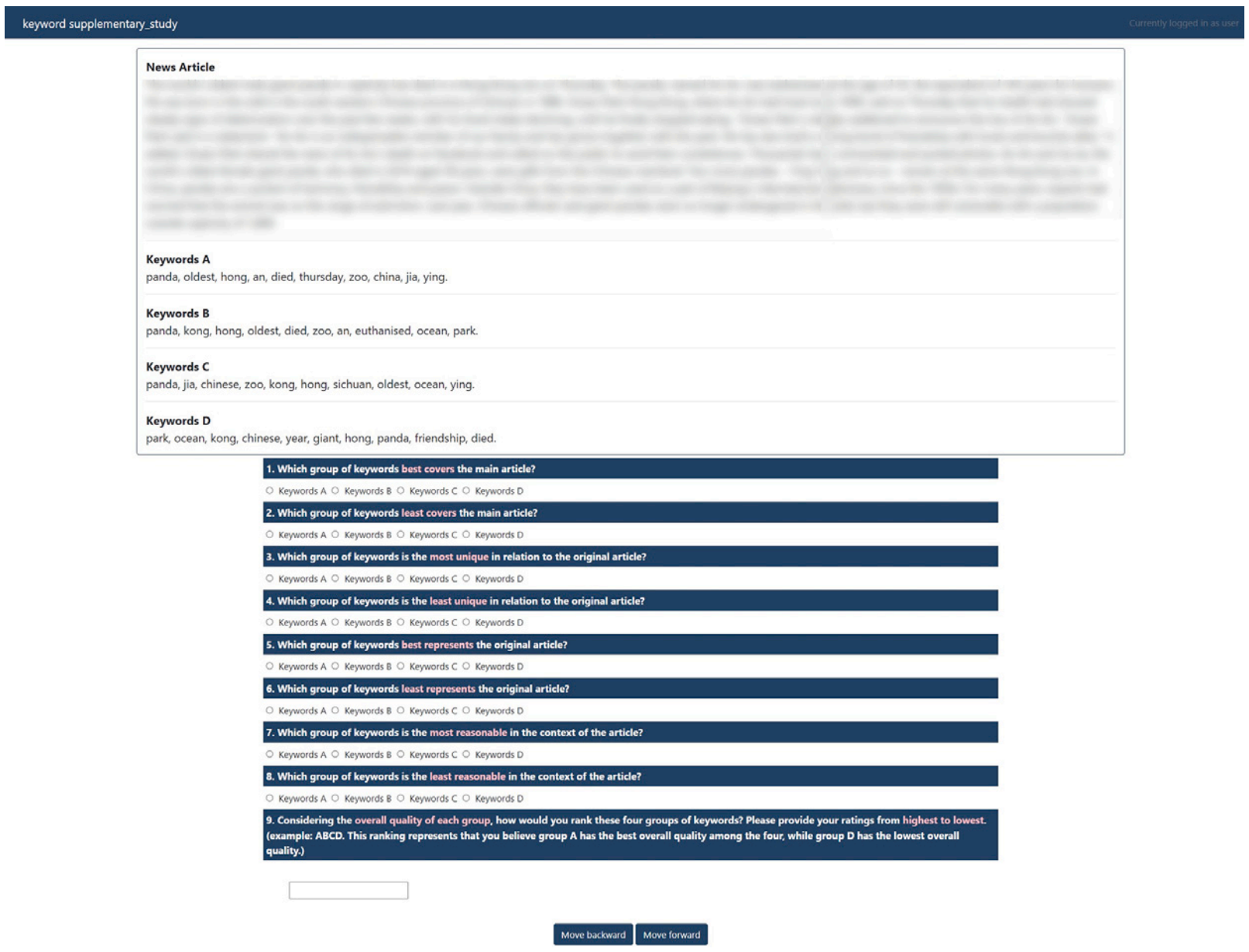

**Fig. E.3.** An example of the task instance page for the supplementary experiment: It contains a base article, four sets of keywords generated by TF-IDF, KeyBERT, Llama 2 and gold standard, 8 single-choice questions and a sequencing question.

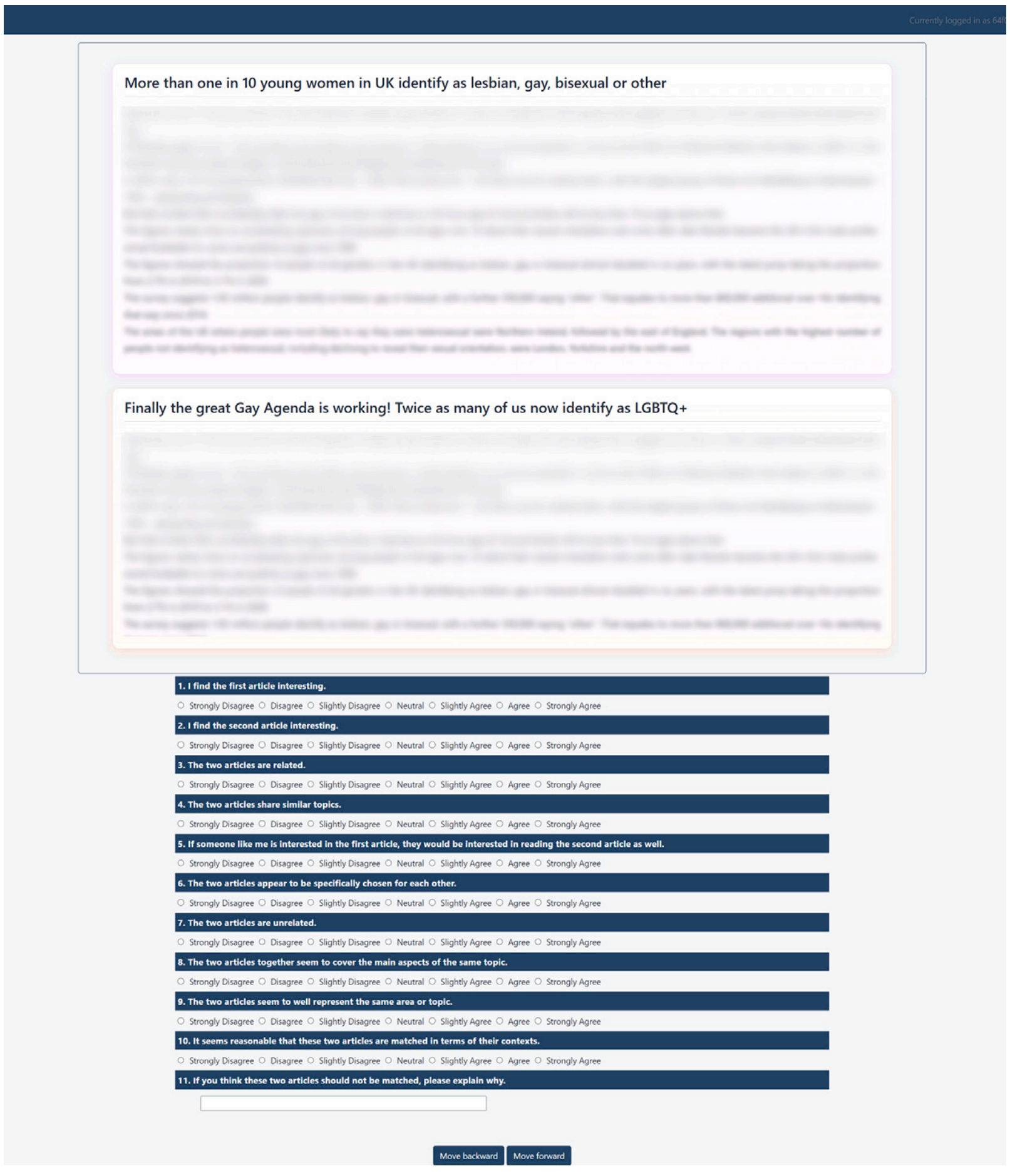

**Fig. E.4.** An example of the task instance page for the downstream experiment: It contains a base article, a matched article, ten 7-point Likert-scale questions, and one optional free-text feedback question.

**Table F.1**
The detailed values of the horizontal analysis boxplot. The values of the mean, the first quartile, the median, the third quartile and the interquartile range ($H_Q3 - H_Q1$) of the standard deviations of all 15 task instances under the alternative phrasings of the questions used are listed.

| Metrics | 1.a | 1.b | 1.c | 2.a | 2.b | 2.c | 3.a | 3.b | 3.c | 4.a | 4.b | 4.c | 5.a | 5.b | 5.c |
|---|---|---|---|---|---|---|---|---|---|---|---|---|---|---|---|
| H_Mean | 1.45 | 1.53 | 1.62 | 1.33 | 1.37 | 1.28 | 1.38 | 1.42 | 1.47 | 1.22 | 1.30 | 1.30 | 1.43 | 1.46 | 1.41 |
| H_Q1 | 1.35 | 1.45 | 1.52 | 1.18 | 1.23 | 1.15 | 1.12 | 1.35 | 1.38 | 1.12 | 1.23 | 1.22 | 1.28 | 1.37 | 1.29 |
| H_Q2 | 1.50 | 1.54 | 1.59 | 1.33 | 1.32 | 1.22 | 1.45 | 1.41 | 1.42 | 1.23 | 1.28 | 1.29 | 1.46 | 1.45 | 1.42 |
| H_Q3 | 1.58 | 1.57 | 1.73 | 1.56 | 1.51 | 1.46 | 1.54 | 1.50 | 1.58 | 1.42 | 1.40 | 1.47 | 1.57 | 1.58 | 1.58 |
| H_IQR | 0.24 | 0.12 | 0.21 | 0.39 | 0.28 | 0.30 | 0.42 | 0.15 | 0.19 | 0.29 | 0.17 | 0.25 | 0.30 | 0.21 | 0.29 |

**Table F.2**
The detailed values of the vertical analysis boxplot. *V_Mean1* represents the value of the mean of the absolute deviations of all participant evaluations under the alternative phrasings of the questions used. *V_Mean2* shows the value of the mean of the standard deviations of these deviations. This table also contains values for the first quartile, the median, the third quartile and the interquartile range ($V_Q3 - V_Q1$) of these standard deviations.

| Metrics | 1.a | 1.b | 1.c | 2.a | 2.b | 2.c | 3.a | 3.b | 3.c | 4.a | 4.b | 4.c | 5.a | 5.b | 5.c |
|---|---|---|---|---|---|---|---|---|---|---|---|---|---|---|---|
| V_Mean1 | 0.41 | 0.42 | 0.41 | 0.28 | 0.31 | 0.29 | 0.36 | 0.36 | 0.39 | 0.36 | 0.45 | 0.38 | 0.32 | 0.32 | 0.30 |
| V_Mean2 | 0.28 | 0.27 | 0.26 | 0.23 | 0.25 | 0.22 | 0.26 | 0.22 | 0.24 | 0.25 | 0.30 | 0.28 | 0.25 | 0.24 | 0.23 |
| V_Q1 | 0.16 | 0.16 | 0.16 | 0 | 0.22 | 0 | 0.16 | 0.16 | 0.16 | 0.22 | 0.22 | 0.22 | 0.16 | 0.16 | 0.16 |
| V_Q2 | 0.27 | 0.27 | 0.27 | 0.22 | 0.22 | 0.22 | 0.22 | 0.16 | 0.16 | 0.22 | 0.22 | 0.22 | 0.22 | 0.22 | 0.22 |
| V_Q3 | 0.33 | 0.32 | 0.31 | 0.38 | 0.38 | 0.38 | 0.33 | 0.31 | 0.31 | 0.38 | 0.38 | 0.38 | 0.33 | 0.38 | 0.31 |
| V_IQR | 0.18 | 0.16 | 0.16 | 0.38 | 0.16 | 0.38 | 0.26 | 0.18 | 0.16 | 0.16 | 0.16 | 0.16 | 0.18 | 0.22 | 0.16 |

**Table F.3**
The table shows the normalised values for different phrasings under each evaluation character across several key metrics. For the results of the horizontal analysis, the phrasing with a lower sum of standard deviations (*H_Mean*) is preferred. If two phrasings have similar sums, then the one with smaller standard deviations is preferred. The third quartile (*H_Q3*) indicates that three-quarters of the data are smaller than this quartile, which is an efficient metric for evaluating the phrasings. Similarly, for the results of the vertical analysis, the mean of the participant evaluations' deviations (*V_Mean1*), followed by the mean of these deviations' standard deviations (*V_Mean2*) and their proportion (*V_Q3*) are prioritised. We then compute a final score for each phrasing taking the aforementioned metrics into account. This final score is primarily determined by the sum of *H_Mean* and *V_Mean1*, plus one-tenth of the sum of the remaining mentioned metrics. The phrasings with the lowest final scores are used in the main experiment.

| Metrics | 1.a | 1.b | 1.c | 2.a | 2.b | 2.c | 3.a | 3.b | 3.c | 4.a | 4.b | 4.c | 5.a | 5.b | 5.c |
|---|---|---|---|---|---|---|---|---|---|---|---|---|---|---|---|
| H_Mean | 0.32 | 0.33 | 0.35 | 0.33 | 0.34 | 0.32 | 0.32 | 0.33 | 0.34 | 0.32 | 0.34 | 0.34 | 0.33 | 0.34 | 0.33 |
| V_Mean1 | 0.33 | 0.34 | 0.33 | 0.32 | 0.35 | 0.33 | 0.32 | 0.32 | 0.35 | 0.30 | 0.38 | 0.32 | 0.34 | 0.34 | 0.32 |
| V_Mean2 | 0.35 | 0.33 | 0.32 | 0.33 | 0.36 | 0.33 | 0.36 | 0.32 | 0.35 | 0.30 | 0.36 | 0.34 | 0.35 | 0.33 | 0.32 |
| H_Q3 | 0.32 | 0.32 | 0.34 | 0.34 | 0.33 | 0.32 | 0.33 | 0.32 | 0.34 | 0.33 | 0.33 | 0.34 | 0.33 | 0.33 | 0.33 |
| V_Q3 | 0.34 | 0.33 | 0.32 | 0.33 | 0.33 | 0.33 | 0.35 | 0.33 | 0.33 | 0.33 | 0.33 | 0.33 | 0.32 | 0.37 | 0.30 |
| Overall | 0.75 | 0.77 | 0.78 | 0.75 | 0.80 | 0.75 | 0.75 | 0.75 | 0.80 | 0.72 | 0.82 | 0.76 | 0.77 | 0.78 | 0.74 |

## Appendix F. Detailed horizontal and vertical analysis

See Tables F.1–F.3.